\documentclass[fleqn,usenatbib]{mnras}

\usepackage{newtxtext,newtxmath}

\usepackage[T1]{fontenc}

\DeclareRobustCommand{\VAN}[3]{#2}
\let\VANthebibliography\thebibliography
\def\thebibliography{\DeclareRobustCommand{\VAN}[3]{##3}\VANthebibliography}

\usepackage{graphicx}	
\usepackage{amsmath}	

\usepackage[textwidth=20mm]{todonotes}
\usepackage{setspace}

\usepackage{dsfont}
\usepackage{tikz}

\graphicspath{{./}{Figures/}}

\newcommand{\bX}{{\boldsymbol{X}}}

\newcommand{\E}{{\mathbb{E}}}

\newcommand{\N}{{\mathcal{N}}}
\newcommand{\J}{{\mathcal{J}}}
\renewcommand{\S}{{\mathcal{S}}}

\newcommand{\R}{{\mathds{R}}}

\newcommand{\M}{{\mathrm{M}}}

\renewcommand{\Finv}{{\breve{F}}}

\title[Age-chemical structure of the Galactic disk]{Decoding the age-chemical structure of the Milky Way disk:\\An application of Copulas and Elicitable Maps}

\author[A. A. Patil et al.]{
Aarya A. Patil$^{1, 2}$\thanks{E-mail: patil@astro.utoronto.ca},
Jo Bovy$^{1}$,
Sebastian Jaimungal$^{3}$,
Neige Frankel$^{4, 1}$,
Henry W. Leung$^{1}$
\\
$^{1}$David A. Dunlap Department of Astronomy \& Astrophysics, University of Toronto, 50 St George Street, Toronto ON M5S 3H4, Canada\\
$^{2}$Dunlap Institute for Astronomy \& Astrophysics, University of Toronto, 50 St George Street, Toronto, ON M5S 3H4, Canada\\
$^{3}$Department of Statistical Sciences, University of Toronto, 9th Floor, Ontario Power Building, 700 University Ave, Toronto, ON M5G 1Z5, Canada\\
$^{4}$Canadian Institute for Theoretical Astrophysics, University of Toronto, 60 St. George Street, Toronto, ON M5S 3H8, Canada
}

\date{Accepted XXX. Received YYY; in original form ZZZ}

\pubyear{2023}

\begin{document}
\label{firstpage}
\pagerange{\pageref{firstpage}--\pageref{lastpage}}
\maketitle

\begin{abstract}
In the Milky Way, the distribution of stars in the $[\alpha/\mathrm{Fe}]$ vs. $[\mathrm{Fe/H}]$ and $[\mathrm{Fe/H}]$ vs. age planes holds essential information about the history of star formation, accretion, and dynamical evolution of the Galactic disk. We investigate these planes by applying novel statistical methods called copulas and elicitable maps to the ages and abundances of red giants in the APOGEE survey. We find that the low- and high-$\alpha$ disk stars have a clean separation in copula space and use this to provide an automated separation of the $\alpha$ sequences using a purely statistical approach. This separation reveals that the high-$\alpha$ disk ends at the same [$\alpha$/Fe] and age at high $[\mathrm{Fe/H}]$ as the low-$[\mathrm{Fe/H}]$ start of the low-$\alpha$ disk, thus supporting a sequential formation scenario for the high- and low-$\alpha$ disks. We then combine copulas with elicitable maps to precisely obtain the correlation between stellar age $\tau$ and metallicity $[\mathrm{Fe/H}]$ conditional on Galactocentric radius $R$ and height $z$ in the range $0 < R < 20$ kpc and $|z| < 2$ kpc. The resulting trends in the age--metallicity correlation with radius, height, and [$\alpha$/Fe] demonstrate a $\approx 0$ correlation wherever kinematically-cold orbits dominate, while the naively-expected negative correlation is present where kinematically-hot orbits dominate. This is consistent with the effects of spiral-driven radial migration, which must be strong enough to completely flatten the age--metallicity structure of the low-$\alpha$ disk. 
\end{abstract}




\section{Introduction}
The formation history of a galaxy is a complex amalgamation of several small and large-scale processes such as star formation in molecular gas clouds, stellar evolution and chemical enrichment of the interstellar medium, galaxy dynamics due to internal and external processes, and pristine gas inflow from the cosmic web \citep{binney_2008}. The Milky Way, a typical disk galaxy in the Universe, can be used as a test bed to disentangle these mechanisms with high-resolution spectroscopic observations on a star-by-star basis \citep{freeman_2002, rix_2013}. Wide-field spectroscopic surveys such as APOGEE \citep{majewski_2017}, Gaia-ESO \citep{gilmore_2012}, GALAH \citep{de_silva_2015}, and the ongoing and future SDSS-V \citep{kollmeier_2017}, WEAVE \citep{dalton_2012}, MOONS \citep{cirasuolo_2014}, and 4MOST \citep{de_jong_2019} surveys allow chemical abundance and age estimates for ${\sim}10^6$ stars in the Galaxy. These estimates thus provide opportunities to rigorously test models of Galaxy formation and evolution.

One of the outstanding questions in Galactic evolution studies is: What is the star formation history and chemical enrichment of the Milky Way disk? Stars inherit the chemical composition of the interstellar medium (ISM) from which they are formed. The chemistry of the ISM evolves over time as stars die and enrich the medium with heavy elements. New stars form in the enriched medium and their cycle of formation, evolution, and death continues. This composite process tells us that the stellar age distribution in the Milky Way disk encodes information about its star formation history and the chemical distribution acts as a fossil record of its enrichment.

The chemical distribution of Galactic disk stars is often investigated using the $[\alpha\mathrm{/Fe}]$ and $[\mathrm{Fe/H}]$ elemental abundances. These abundances are key diagnostics of enrichment as Type II (core collapse) and Type Ia supernovae predominantly producing $\alpha$ and iron peak elements respectively. It has long been observed that stars across the Galactic disk exhibit a bimodality in the $[\alpha\mathrm{/Fe}]-[\mathrm{Fe/H}]$ space \citep[e.g.,][]{fuhrmann_1998, bensby_2003, anders_2014, nidever_2014, hayden_2015, kordopatis_2015}. The disk is thus believed to have two components, the high- and low-$\alpha$ sequences. Previously, these two sequences were represented in terms of their spatial distributions as the geometrically defined thick and thin disks \citep{yoshi_1982, gilmore_1983}. The thick disk generally corresponds to the $\alpha$-enhanced population of old, kinematically hot stars whereas the thin disk to the $\alpha$-poor population of young, kinematically cold stars \citep[e.g.,][]{fuhrmann_1998, feltzing_2003}. However, the scale length and height of the chemically-separated components do not have a one-to-one correspondence with those of the geometric thin and thick disks \citep{schonrich_binney_2009b, bovy_2012a, bovy_2012b, bovy_2012c, martig_2016, bovy_2016, bland-hawthorn_2019}. The decomposition is now better understood in terms of the chemical ($\alpha$) enrichment, as studied using high-resolution spectroscopic surveys such as APOGEE \citep[e.g.,][]{anders_2014}. However, the separation between the $\alpha$ sequences \citep{nidever_2014} as well as the mere existence of these sequences \citep[e.g., Gaia-ESO results in][]{recio-blanco_2014, rojas_2014} is still a matter of debate.

Observational tests suggest that the sequences are distinct populations \citep{hayden_2015, mackereth_2019_dynamic, gandhi_2019} with each sequence having a different timescale of formation. While an inside-out disk formation \citep[e.g., the two-infall model in][]{chiappini_2001} and its upside-down evolution \citep{stinson_2012, bird_2013} explains the presence of these distinct populations, other formation scenarios invoking different external and internal processes such as disk heating by minor mergers \citep{quinn_1993, villalobos_2008}, stellar accretion of satellites \citep{abadi_2003}, early gas-rich mergers \citep{brook_2004} or massive accretion events \citep{belokurov_2018, helmi_2018}, star formation in gas-rich clumps at high redshift \citep{clarke_2019}, and radial migration \citep{schonrich_binney_2009a, loebman_2011, sharma_2021} are plausible. Particularly, the in-situ scenario in \citet{schonrich_binney_2009a} (see also \citealt{sharma_2021}) allows for a continuous star formation history through the secular process of radial migration, which is expected to be ubiquitous in spiral disks. However, it is unclear if radial migration alone can lead to the thickening of the high-$\alpha$ disk \citep{minchev_2012, kawata_2016} and whether the $\alpha$ bimodality is found across disk galaxies \citep[e.g.,][]{mackereth_2018}. Simulations in \cite{buck_2019} (see also \citealt{buck_2023}) demonstrate that a combination of gas-rich merger and radial migration could also explain the bimodality, but we need to dissect the observed $\alpha$ sequences using sophisticated statistical approaches to constrain the effects of individual processes underlying the Milky Way.

In this paper, we present the first automated approach to separate the $\alpha$ sequences using a powerful statistical tool called copulas. Particularly, we apply copulas to the $[\alpha\mathrm{/Fe}]-[\mathrm{Fe/H}]$ plane of the APOGEE DR17 stars in the Milky Way disk. Copulas extract the dependence structure of random variables without being affected by their marginal distributions \citep{nelsen_2006}. Thus, we can understand how $[\alpha\mathrm{/Fe}]$ correlates with $[\mathrm{Fe/H}]$ across the disk irrespective of its overall chemical distribution. 

In addition to chemical abundances, ages of stars provide a detailed view into the evolutionary history of the Galaxy. Previously, one zone models of Galactic chemo-dynamical evolution predicted a tight correlation between the ages and metallicities of stars at any given radius of the Milky Way. However, \cite{edvardsson_1993} and several others identified a spread in the age-metallicity (age-[Fe/H]) relation in the solar neighbourhood, and recent studies have validated this beyond the immediate solar neighborhood of the Galactic disk \citep{feuillet_2019}. Despite using different target stellar types and age estimation methods, the results consistently show that stars of a given metallicity cover a wide range of ages and the most metal-rich stars have intermediate ages. Small sample studies with small age uncertainties show that the age-metallicity spread in the solar neighbourhood is intrinsically present and not due to observational errors \citep[e.g.][]{bensby_2014, nissen_2018}.

The most successful to explain the absence of a clear stellar age-metallicity relation is radial migration: the process of changing stars' guiding-center radii without significantly increasing their eccentricity. The archetypical process of this type is churning at corotation by transient spiral structure \citep{sellwoodbinney2002}, but other dynamical processes can give rise to a similar behavior \citep{minchev_2010}. Radial migration is backed by both analytical models and simulations \citep{sellwoodbinney2002, roskar_2008, minchev_2010} and it can reproduce much of the chemical behavior of the low-$\alpha$ disk \citep[e.g.,][]{hayden_2015, frankel_2018, feuillet_2019}. In particular, we know that the metallicity of the ISM decreases outwards in the disk \citep{schonrich_binney_2009b}, and this metallicity gradient along with stellar migration can explain the age-metallicity spread in the disk.  

There is a growing need to confirm the spread in age-metallicity and assess radial migration throughout the Milky Way disk. \cite{frankel_2018} test radial migration more ``globally" by measuring its efficiency using a dataset of red clump stars over a Galactocentric radius range of $5 \leq R \leq 14$ kpc. They include the effects of radial heating and radial migration into a \textit{radius migration} model, and constrain the migration strength of a typical star over the disk lifetime (${\sim}$8 Gyr) to $\sigma_\mathrm{RM8} = 3.6 \sqrt{\tau/8}$ kpc where $\tau$ is the stellar age in Gyr. \cite{frankel_2020} improve this model by disentangling the strength of radial heating and radial migration, but the model is still fitted to a limited portion of the disk. In this paper, we investigate the stellar age-metallicity structure over a large fraction of the Milky Way disk by combining copulas with another novel statistical method: elicitable maps. These maps in combination with neural networks allow us to estimate continuous functions of random variables given sparse observed data; we use them to estimate the correlation between stellar age and metallicity across the disk.

One of the aims of this paper is to demonstrate that copulas and elicitable maps are powerful statistical tools with potential applications across different subfields in astronomy. The concept of copulas is fundamental to statistics and probability theory, and its applications have been influential in the field of mathematical finance and in several areas of engineering; they have only been applied sparingly in astronomy \citep[e.g., cosmology applications in][]{scherrer_2009, sato_2010, benabed_2009, vio_2020} even though they provide non-linear measures of dependence and other advantages as compared to the commonly-used Pearson correlation. On the other hand, the term elicitability was recently coined in the statistics literature \citep{lambert_2008} although its principles have been around since \cite{osband_1985}. This method has no known usage in astronomy to the best of our knowledge, but has tremendous potential for astronomical analysis in the presence of sparse data. To aid with applications of copulas and elicitable maps, we are developing a generalized Python package available online at \url{https://github.com/aaryapatil/elicit-disk-copulas}.

The outline of the paper is as follows. We describe the APOGEE spectroscopic data we use in this study in Section \ref{sec:data}. Section \ref{sec:copulas} introduces copulas and explains how they extract dependence structures of random variables. In Section \ref{subsec:septracks}, we apply copulas to automatically separate the bimodal distribution of Galactic disk stars in $[\alpha\mathrm{/Fe}]-[\mathrm{Fe/H}]$ space. We then discuss elicitable maps in Section \ref{sec:elicitable} and combine them with copulas to obtain the correlation between age and metallicity of disk stars as a continuous function of Galactocentric radius $R$ and vertical distance from the midplane $z$. Section \ref{subsec:cond_corr_cop} provides a detailed explanation of this procedure and shows the spatial distribution of the correlation between stellar age and metallicity for the low- and high-$\alpha$ sequences. In Section \ref{sec:results}, we discuss our split of the $\alpha$ sequences and estimate of age-metallicity correlation across the two-component disk. The implications of these results on stellar radial migration models in the Galactic disk are also discussed in this section. In the concluding Section \ref{sec:conclusion}, we summarize this paper and its prospects for future studies.

\begin{figure}
    \centering
    \includegraphics[width=1.0\linewidth]{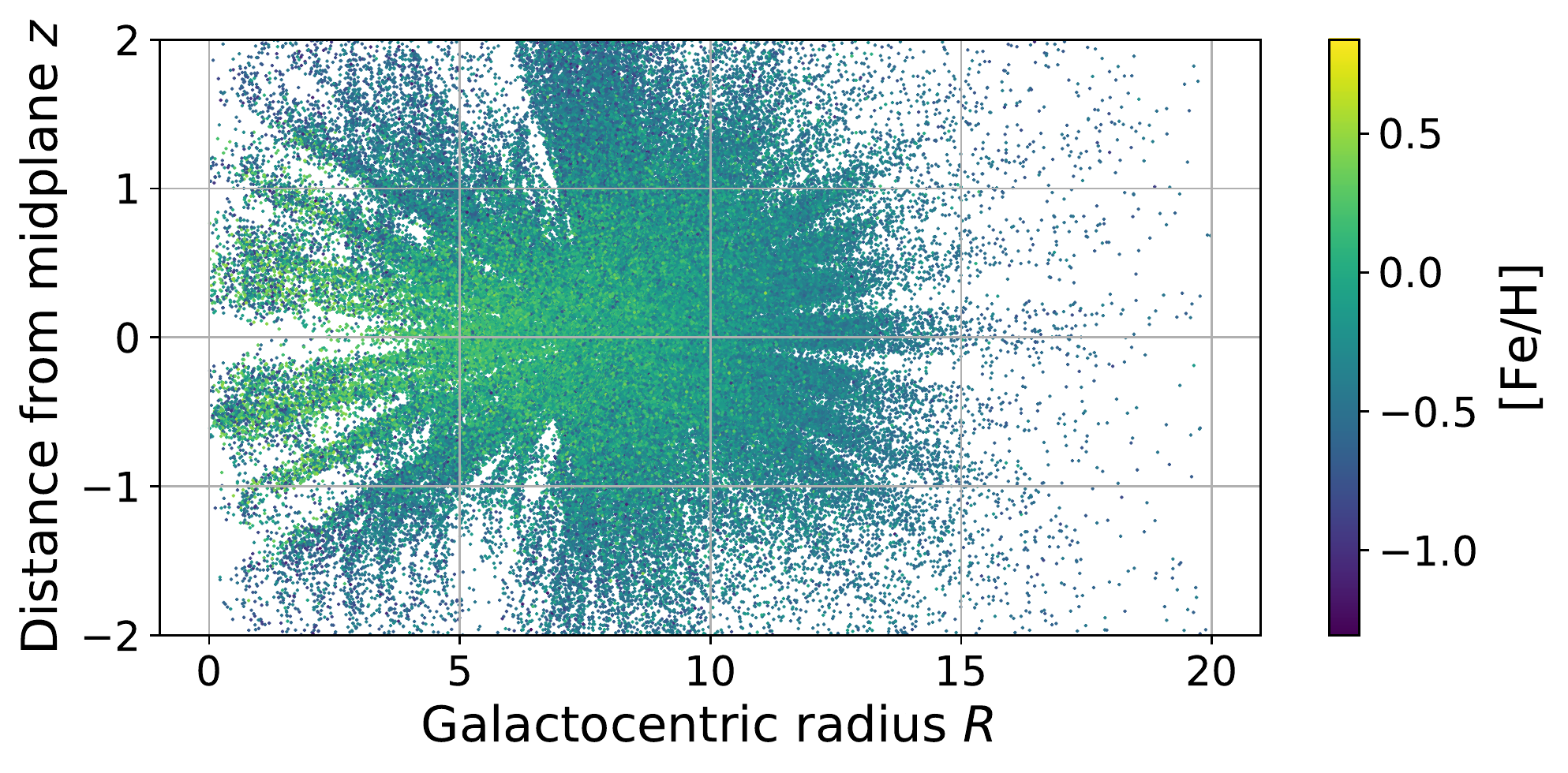}
    \caption{Spatial distribution in Galactocentric radius $R$ and distance from the midplane $z$ of our APOGEE DR17 sample of stars and their metallicities $[\mathrm{Fe/H}]$. We obtain this subset using the quality and selection cuts described in Section \ref{sec:data}.}
    \label{fig:stellar_sample_space}
\end{figure}

\begin{figure*}
    \centering
    \includegraphics[width=1.0\linewidth]{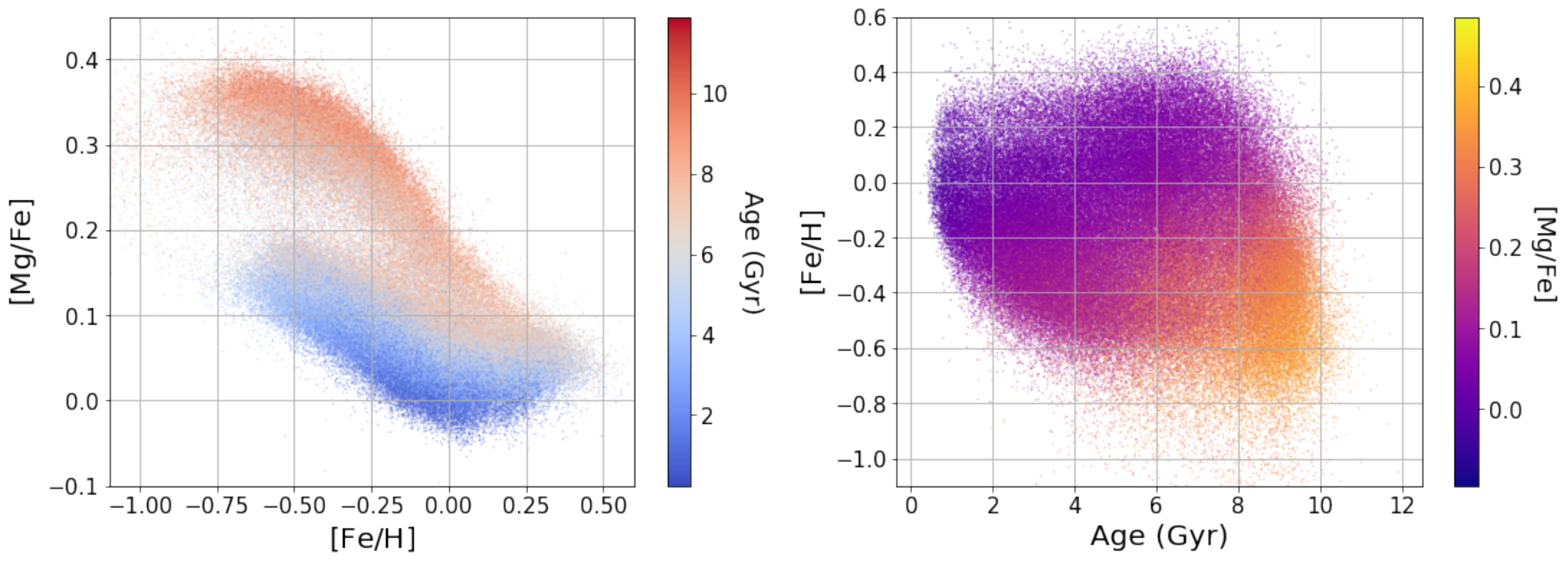}
    \caption{Our APOGEE DR17 stellar sample in the $[\mathrm{Fe/H}]$-$[\alpha\mathrm{/Fe}]$ (left) and age-$[\mathrm{Fe/H}]$ (right) space color-coded by age and $[\alpha\mathrm{/Fe}]$ respectively. These estimates are obtained from the \texttt{astroNN} value-added catalog.}
    \label{fig:stellar_sample}
\end{figure*}

\begin{figure*}
    \centering
    \includegraphics[width=1.0\linewidth]{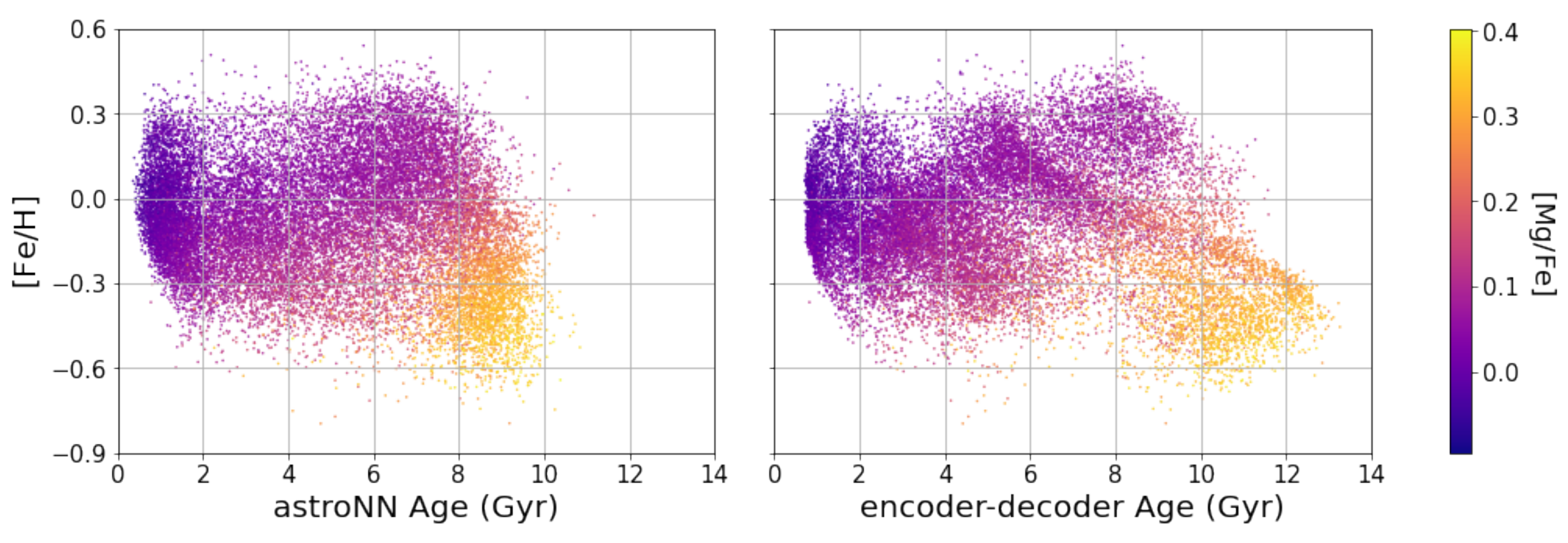}
    \caption{APOGEE stellar samples used in this study. The left panel shows the age-metallicity structure of our \emph{updated} sample using \texttt{astroNN} age estimates, whereas the right panel shows this structure using more accurate ages \citep{leung_2023}.}
    \label{fig:leung_data}
\end{figure*}

\section{Data}\label{sec:data}
Our spectroscopic sample of stars comes from Data Release 17 \citep[DR17;][]{abdurro'uf_2022} of the Sloan Digital Sky Survey-IV \citep[SDSS-IV;][]{blanton_2017} Apache Point Observatory Galactic Evolution Experiment \citep[APOGEE;][]{majewski_2017}. APOGEE is a high-resolution H-band spectroscopic survey of stars that are primarily in the Milky Way. The APOGEE spectrographs \citep{wilson_2019} on the Sloan Foundation \citep{gunn_2006} and du Pont \citep{bowen_1973} 2.5m telescopes are equipped with 300 fibers, each capable of observing a spectrum ranging from 1.51 to 1.69 $\mu m$ with a resolution R $\sim$ 22500. APOGEE DR17 has a total of 2,659,178 spectra reduced using the pipeline in J. Holtzman et al. 2022, in preparation \citep{nidever_2015}.

We use stellar parameter and abundance estimates in the astroNN value-added catalog of DR17. \cite{leung_2019} obtain these estimates and their uncertainties by applying an efficient deep neural network to the APOGEE DR17 stellar spectra. The astroNN catalog also computes distance \citep{leung_2019b} and age \citep{mackereth_2019_dynamic} estimates of DR17 stellar spectra by training on Gaia EDR3 \citep{gaia_2021} and APOKASC-2 \citep{pinsonneault_2018} respectively, which we use in our study.

We construct our stellar sample in two steps. First, we apply the following quality cuts to obtain a subset of stars observed in APOGEE DR17:
\begin{enumerate}
    \item Select stars with tag \texttt{EXTRATARG==0}: These stars represent the main survey (red star) targets. We refer to \citet{zasowski_2013} and \citet{zasowski_2017} for details of the APOGEE target selection.
    \item Remove stars whose \texttt{STAR\_BAD} or \texttt{SN\_BAD} bits in the \texttt{ASPCAPFLAG} bitmask are set: These stars either have ASPCAP fit issues or low signal-to-noise ratio ($S/N < 50$)
    \item Remove stars whose \texttt{BAD\_PIXELS} or \texttt{VERY\_BRIGHT\_NEIGHBOR} bits in the \texttt{STARFLAG} bitmask are set: These stars have pixel level issues or a bright neighbour that affects its spectral observation.
    \item Remove stars with tag \texttt{MEMBER==1}: These stars are in known star clusters or nearby dwarf galaxy members
\end{enumerate}

We then apply the following selection criteria using estimates of effective temperature $T_\mathrm{eff}$, surface gravity $\log g$, metallicity $[\mathrm{Fe/H}]$, age $\tau$, distance $d$ and its uncertainty $\Delta d$, Galactocentric radius $R$, distance from the midplane $z$ in the \texttt{astroNN} value-added catalog:
\begin{enumerate}
    \item $4000 \, \mathrm{K} < T_\mathrm{eff} < 5500 \, \mathrm{K}$
    \item $1.0 < \log g < 3.8$
    \item $[\mathrm{Fe/H}] > -2$
    \item $\tau < 12$ Gyr
    \item $\frac{\Delta d}{d} < 0.2$
    \item $R < 20$ kpc
    \item $\lvert z \rvert < 2$ kpc.
\end{enumerate} Note that \texttt{astroNN} computes $R$ and $z$ using the solar position of $R = 8.125$ kpc \citep{gravity_2018} and $z = 20.8$ pc \citep{bennett_2019}. The final sample is representative of the Milky Way disk and includes a total of 182,570 APOGEE stellar spectra. We refer to this as our \emph{original} sample. Figure \ref{fig:stellar_sample_space} shows the distribution of our stellar sample in Galactocentric radius $R$ and distance from the midplane $z$.

In Figure \ref{fig:stellar_sample}, we show the age-$[\mathrm{Fe/H}]$-$[\alpha\mathrm{/Fe}]$ space of this sample. 
We use magnesium as a representative element of $\alpha$-related enrichment processes like core collapse supernovae. This figure shows that the \texttt{astroNN} stellar ages of our sample are limited to ${\sim}$10 Gyrs even though we expect the high-$[\alpha\mathrm{/Fe}]$ stars to be older. For investigating the age-metallicity relation, we thus use a more recent APOGEE DR17 age catalog that provides more accurate spectroscopic ages than \texttt{astroNN} using a variational encoder-decoder \citep{leung_2023}. The new age catalog includes ages for only a portion of our sample because the encoder-decoder model is applicable to $ 2.5 < \log g < 3.6$ (limited by the training set). We apply this $\log g$ cut along with other quality cuts recommended in \cite{leung_2023} on top of the selection criteria mentioned above. These additional cuts are described as follows
\begin{enumerate}
    \item \texttt{ASPCAPFLAG==0}
    \item \texttt{STARFLAG==0}
    \item latent-space age uncertainty $\sigma_{\tau}$ < 40\%
    \item $\log g > 2.55$
\end{enumerate} Note that the additional $\log g > 2.55$ limit helps avoid the edge of the model parameter space and successfully removes some of the problematic young high-$[\alpha\mathrm{/Fe}]$ stars in the age catalog. These cuts result in a sub-sample of 22,527 stars (referred to as our \emph{updated} sample) whose \cite{leung_2023} ages are compared with those from \texttt{astroNN} in Figure \ref{fig:leung_data}. We see that the high-$[\alpha\mathrm{/Fe}]$ stars in the new catalog are much older (as expected) than those in \texttt{astroNN}, which is a result of improvement in age precision to ${\sim}22$\% precision using the variational encoder-decoder approach.

\section{Copulas}\label{sec:copulas}
In this section, we introduce copulas as a way to study the dependency structure of chemical and age data and then use it to study the $[\alpha\mathrm{/Fe}]-[\mathrm{Fe/H}]$ plane.  Section \ref{subsec:math_cop} lays down the mathematical foundation of distribution functions called copulas that encapsulate the dependence structures of random variables. Section \ref{subsec:septracks} shows that copulas allow us to better understand the relation between $[\alpha/\mathrm{Fe}]$ and $[\mathrm{Fe/H}]$ of stars, and we develop an approach for automatically separating the low- and high-$[\alpha/\mathrm{Fe}]$ sequences, i.e., the $[\alpha/\mathrm{Fe}]$ bimodality.

\subsection{Mathematical background}\label{subsec:math_cop}
\begin{figure*}
    \centering
    \includegraphics[width=1.0\linewidth]{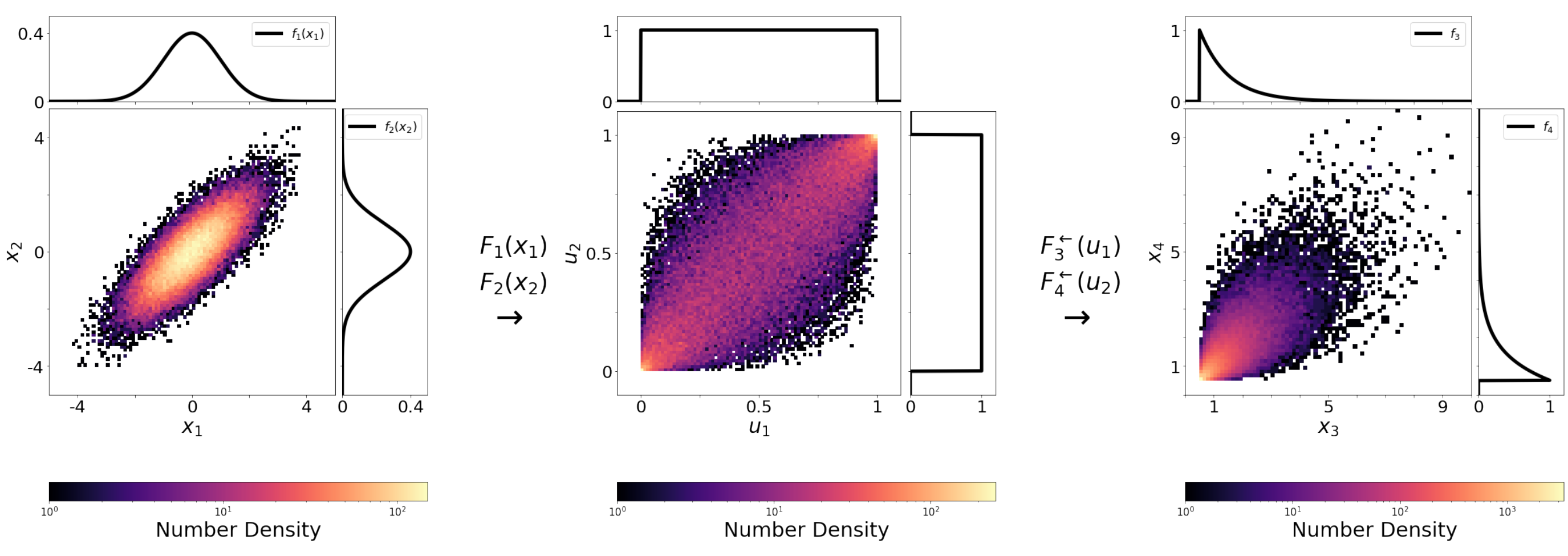}
    \caption{Samples from a two-dimensional Gaussian distribution (left panel) and its corresponding copula, also called the Gaussian copula (middle panel). The right panel shows the sampling of a two-dimensional distribution with the same Gaussian copula but different (exponential) margins.}
    \label{fig:gauss-copula}
\end{figure*}

A copula is the distribution function one obtains upon transforming each variable in a $d$-dimensional data set by applying its cumulative distribution. 
The resulting distribution function has standard uniform margins; thus, the copula captures the dependency structure of variables without being affected by their marginal distributions. In order to provide a more intuitive explanation, we present a copula example in Figure \ref{fig:gauss-copula}. The left panel shows samples $x_1, x_2$ from a two-dimensional Gaussian distribution and its Gaussian margins $f_1, f_2$. The middle panel shows the transformation of the data samples to copula space $u_1, u_2$ by applying the cumulative distribution functions (CDFs) $F_1, F_2$ to variables $x_1, x_2$ respectively. The scatter plot in this panel represents samples from a copula, which in this case it is called a Gaussian copula. We can see that it satisfies the required property of having standard uniform margins. The right panel of the figure shows samples of a distribution that combines the Gaussian copula with exponential margins. To obtain these new samples, we transform $u_1, u_2$ using the inverse of the CDFs $F_3, F_4$ of two exponential distributions (details of this transformation are deferred until later in this section). Thus, as the figure shows, despite having the same underlying copula, the two-dimensional distribution on the left seems to have a symmetric dependency structure whereas the two-dimensional distribution on the right appears to be skewed towards lower values. These two distributions look vastly different solely due to their marginal distributions, despite their dependence structure being identical. This illustrates the power of copulas to extract the pure dependence between the variables in a manner that is unaffected by marginals.

\begin{figure*}
    \centering
    \includegraphics[width=0.9\linewidth]{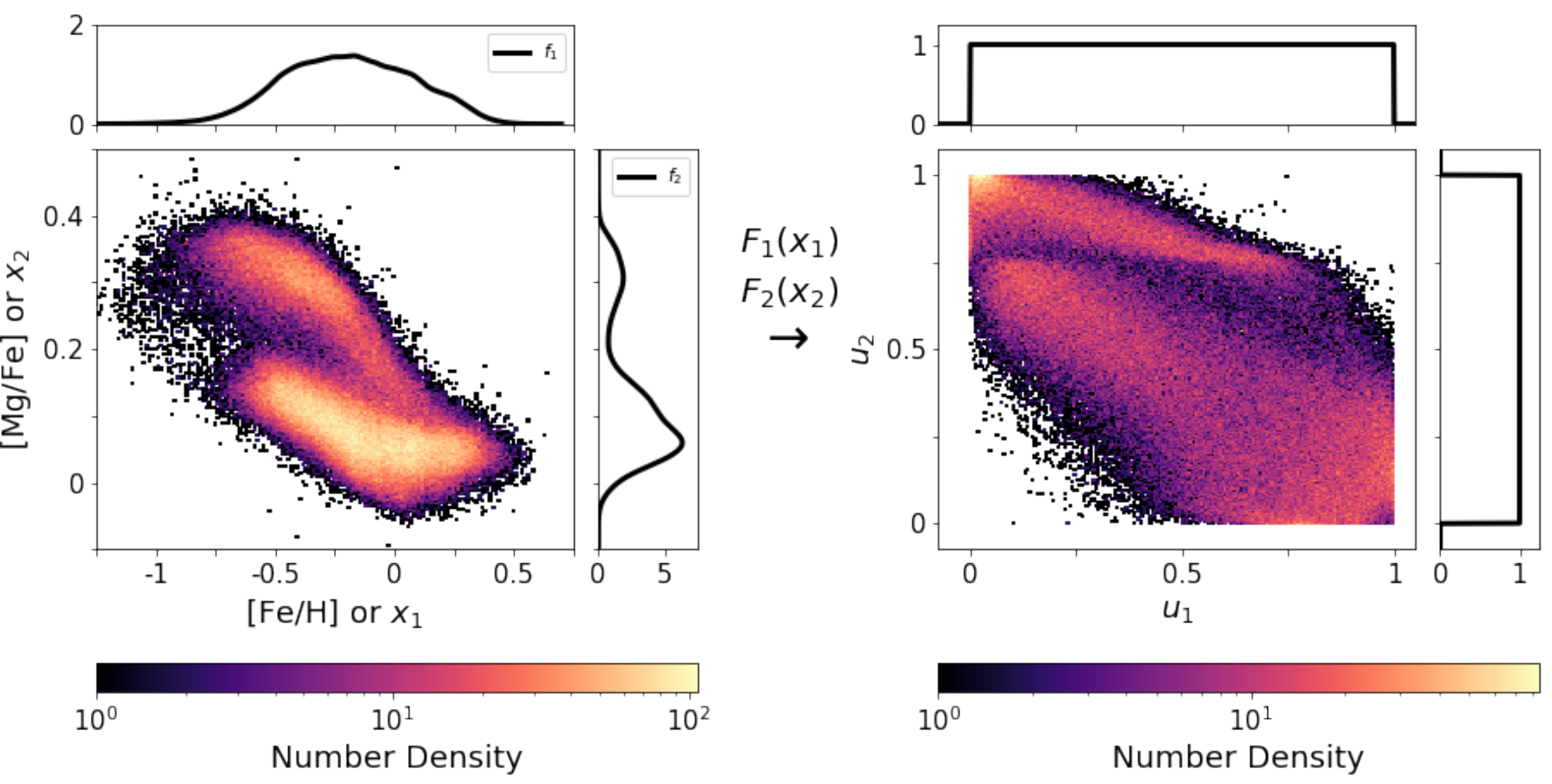}
    \caption{The $[\mathrm{Mg/Fe}]-[\mathrm{Fe/H}]$ space of our \emph{original} APOGEE DR17 sample and its (probability) transformation to copula space. Here $f_1, f_2$ and $F_1, F_2$ are probability distributions and cumulative distribution functions of $x_1, x_2$ or $[\mathrm{Fe/H}], [\mathrm{Mg/Fe}]$ respectively. $u_1, u_2$ represent uniformly distributed variables.}
    \label{fig:MgFe_copula}
\end{figure*}

More formally, a copula is defined as a multivariate CDF $C$ with uniform marginals on the standardized interval $\mathbf{I} =[0, 1]$. In particular, a $d$-dimensional copula is given by $C:\mathbf{I}^d \to \mathbf{I}$. Note that the CDF is also called the distribution function for short and both these terms are interchangeably used. We refer the reader to \cite{nelsen_2006} for a more comprehensive introduction to copulas using statistics and probability theory.

Sklar's theorem \citep{sklar_1959} serves as a foundation for the theory of copulas. First, we need to two definitions
\begin{enumerate}
    \item Probability transformation: for a univariate random variable $X$ having a CDF $F$, $U = F(X)$ is uniform on $\mathbf{I}$.
    \item Quantile transformation: which is the (left) inverse of the probability transformation, i.e., $X = \Finv(U)$.
\end{enumerate}
Next, Sklar's theorem states that, for any multivariate CDF $F$, there exists a copula $C$ such that
\begin{equation}\label{eq:sklar}
    F(x_1, \dotsc, x_d) = C(F_1(x_1), \dotsc, F_d(x_d)), \; \mathbf{x} \in \R^d
\end{equation} 
where $F_1, \dotsc, F_d$ are the one-dimensional marginals of $F$. Moreover, $C$ may be obtained from $F$ as follows
\begin{equation}
C(u_1, \dotsc, u_d) = F(\Finv_1(u_1), \dotsc, \Finv_d(u_d))
\end{equation} 

The converse of the Sklar's theorem also holds: Given any copula $C$ and univariate CDFs $F_1, \dotsc, F_d$, then $F$ defined in \eqref{eq:sklar} is a multivariate distribution function with margins $F_1, \dotsc, F_d$.

\begin{figure*}
    \centering
    \includegraphics[width=0.9\linewidth]{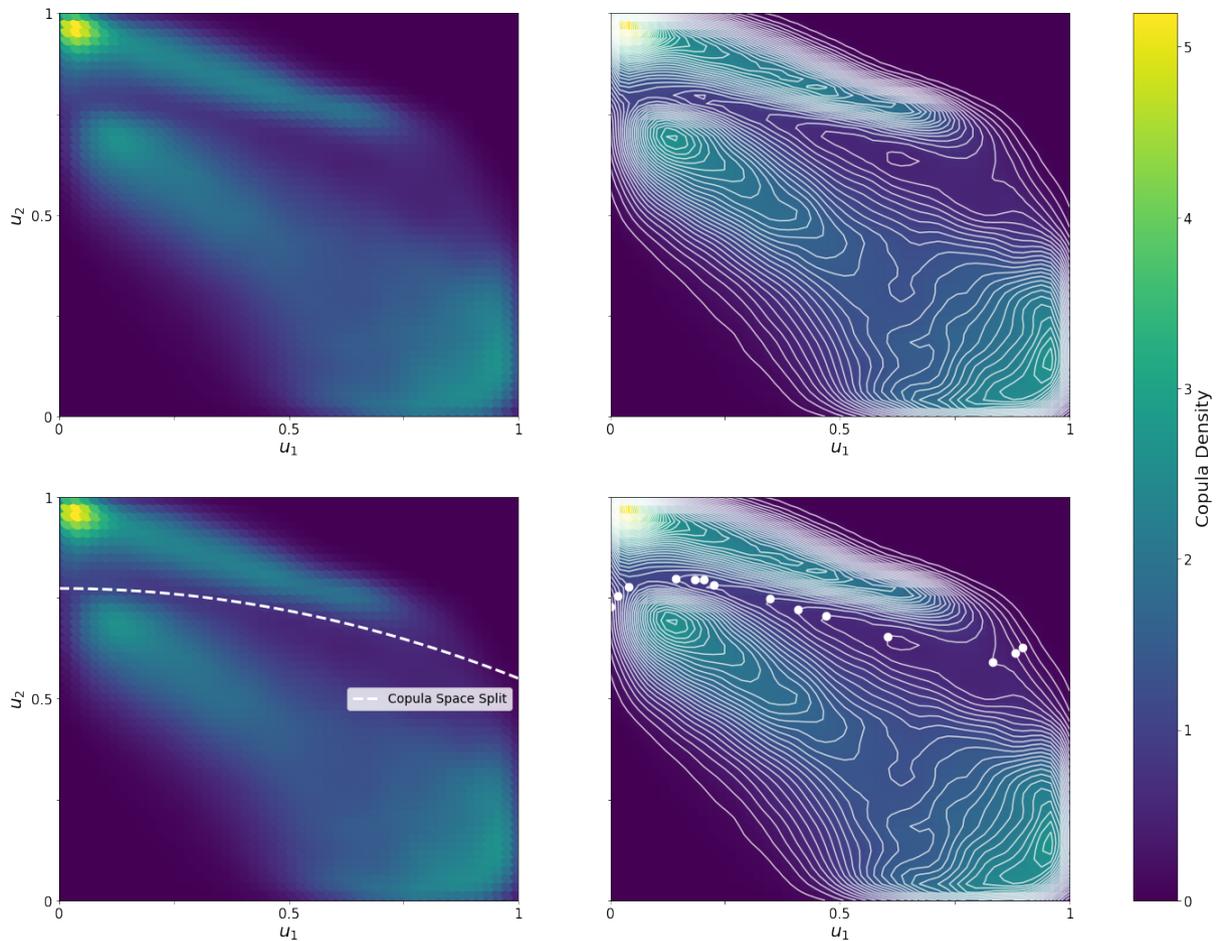}
    \caption{Schematic diagram illustrating our automated procedure for splitting the high- and low-$\alpha$ sequences in the $[\mathrm{Mg/Fe}]-[\mathrm{Fe/H}]$ copula space.}
    \label{fig:split_copula}
\end{figure*}

An important advantage of using copulas is that they follow the invariance principle. Copulas are said to couple $d$-dimensional CDFs to their one-dimensional margins and provide ``scale-invariant” measures of dependence, i.e., they are invariant under applications of strictly increasing transformations. For example, if $C$ is a two-dimensional copula of the distribution function $F(x_1, x_2)$, it is also a copula of $F(g(x_1), g(x_2))$, where $g(x)$ is a strictly increasing function such as $\log(x)$. Indeed, this generalizes to the case when each coordinate is transformed by different increasing functions. In contrast, linear correlations are scale-dependent and cannot properly describe non-linear dependence. Thus, estimating copulas and margins separately allows accurate modelling of multivariate distributions.

The invariance principle is particularly relevant for our applications, where $[\mathrm{Fe/H}]$ and $[\mathrm{Mg/Fe}]$ are logarithmically scaled measurements of chemical abundances but it is unclear if that is the right space to look for correlations in. This property of copulas means that our results are independent of the choice of $Z/Z_\odot$ or $[\mathrm{Fe/H}]$ or similar for the abundance ratio.

\subsection{Separation of low- and high-$\alpha$ disk stars}\label{subsec:septracks}
Two of the advantages of using copulas are that they allow us to analyse the dependency between variables independent of margins and in a scale-invariant way. This means that we can look at the relation between $[\mathrm{Mg/Fe}]$ and $[\mathrm{Fe/H}]$ of stars in the MW disk without being affected by the star formation history (the margins). Thus, we compute the copula of the $[\mathrm{Mg/Fe}]-[\mathrm{Fe/H}]$ distribution of our stellar sample in Figure \ref{fig:stellar_sample}, and display it in Figure \ref{fig:MgFe_copula}. The right panel of this figure shows the transformation of the data in copula space using the probability transformation. Note that we use our \emph{original} sample here because we are looking at the dependency of abundances in this section and do not use ages. Note also that we simply compute the copula of the entire sample rather than splitting it into Galactic spatial zones, because it is well established that the location of the sequences is independent of position in the Galaxy (e.g., \citealt{hayden_2015, weinberg_2019}), even though the distribution of stars along the sequences changes with position, but the latter is ignored by the copula.

Figure \ref{fig:MgFe_copula} shows that the bimodal distribution of $[\mathrm{Mg/Fe}]-[\mathrm{Fe/H}]$ has a cleaner separation in the copula space as opposed to the data space. The high-$\alpha$ sequence in this bimodality is representative of an old population with rapid enrichment whereas the low-$\alpha$ stars are younger and expected to have a gradual enrichment history. The standard way of separating the two sequences is to manually draw a line or polynomial in the data space \citep[for example in][]{mackereth_2019_accretion, weinberg_2019}. We instead use the copula space to automatically separate the sequences. 

\begin{figure*}
    \includegraphics[width=0.8\linewidth]{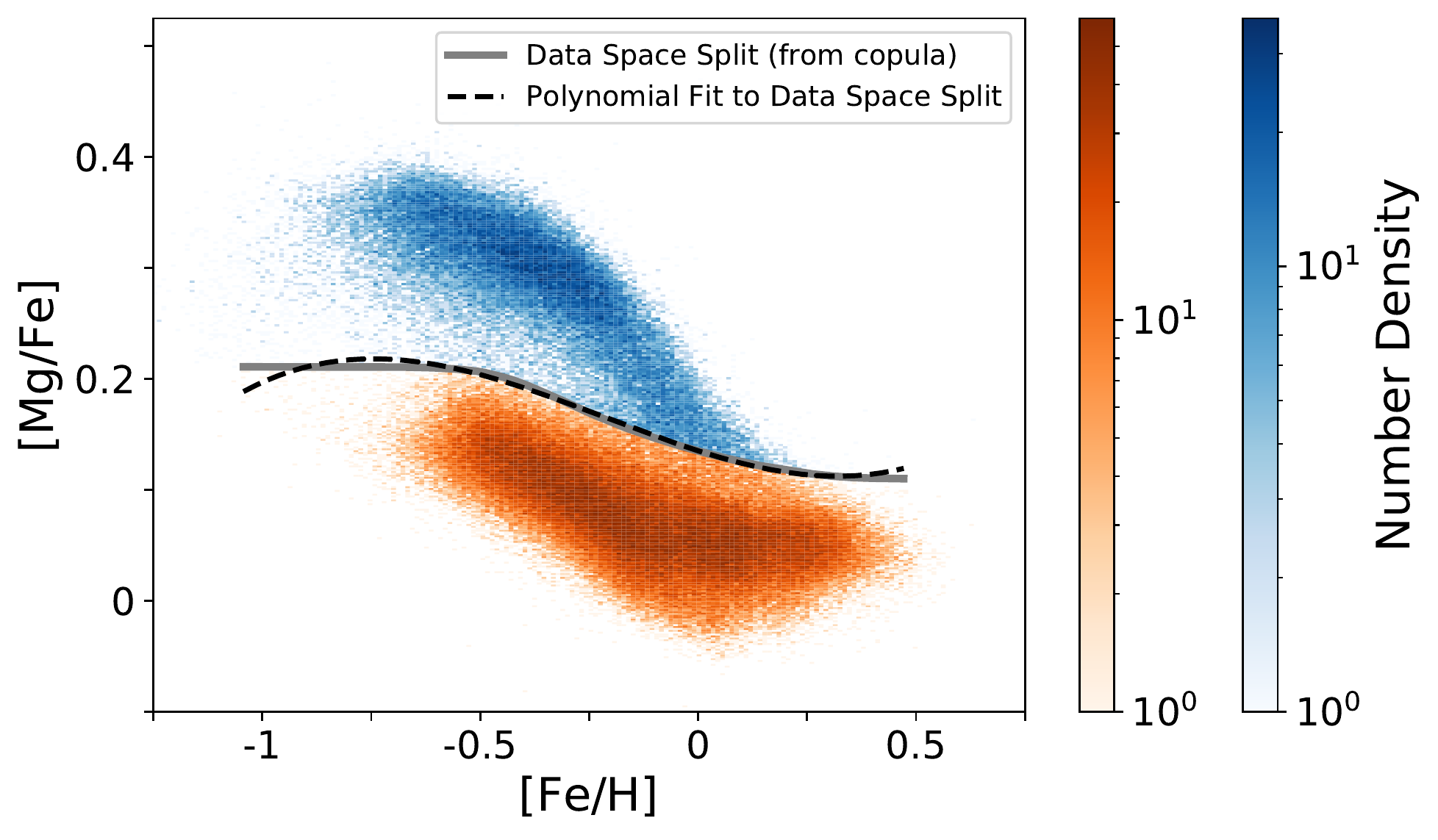}
    \caption{The high- and low-$\alpha$ sequences in data space corresponding to the split in copula space as shown in Figure \ref{fig:split_copula}. The grey line is the data space split, i.e., the copula space split transformed to data space. The black dotted line is a third-degree polynomial fit to this data space split (see Equation \ref{eq:poly_fit_copula}).  Separating the high- and low-$\alpha$ sequences in copula space, it becomes clear that the high-$\alpha$ sequence ends at high $[\mathrm{Fe/H}]$ at the same $[\mathrm{Mg/Fe}]$ where the low-$\alpha$ sequence starts at low $[\mathrm{Fe/H}]$.}
    \label{fig:split_data}
\end{figure*}

The schematic diagram in Figure \ref{fig:split_copula} shows our automated procedure for splitting the high- and low-$\alpha$ sequences. First, we obtain the copula $C$ density corresponding to the copula samples in the right panel of Figure \ref{fig:MgFe_copula}. To do this, we compute a Gaussian kernel density estimate (KDE) approximation to the copula samples along a set of grid points spanning $I^2$. We show this KDE estimate of copula density in the top left panel of Figure \ref{fig:split_copula}. The estimated density shows a bimodal distribution corresponding to the high- and low-$\alpha$ sequences. To automatically separate the sequences in copula density, we plot 54 contour levels between the maximum and minium density values using the \texttt{contour} subpackage in \texttt{matplotlib}. In the right panels of Figure \ref{fig:split_copula}, we show the contours as well as the points on these contours that separate the high copula density regions corresponding to the two sequences. One can think of these ``separating" contour points as a flow line in a vector field of copula density gradient. We explain this as follows.

One can think of the copula density as a topographic map, where the contours represent increasing elevation. Then the flow line we are looking for is the pathway followed by a river between the two elevated terrains. This flow line can be numerically estimated by computing the partial derivatives of the copula density at grid points $u_1, u_2 \in [0, 1]$, which represents a vector field of copula density gradient. However, due to the complexity of this estimation procedure, we use the visual representation of the contours lines to manually choose points where the two elevated terrains meet. Finally, we fit a second-degree polynomial to these points. We show this polynomial in the bottom left panel of Figure \ref{fig:split_copula}.

The above described polynomial curve splits the high- and low-$\alpha$ sequences in copula space. We now transform this curve back to data (real) space using quantile transformation and show this data space split and the resulting two sequences in Figure \ref{fig:split_data}. This split shown in grey when expressed as a polynomial requires six or more degrees, i.e., it does not have a simple polynomial form. However, to make the split accessible for future use, we approximate it using a third-degree polynomial. In other words, the third-degree polynomial is a fit, in data space, to the result obtained in copula space with a second-degree polynomial. This data space fit is shown in Figure \ref{fig:split_data} as a black dashed curve, and we find that it is a good approximation to the actual split in grey. The following equation provides the coefficients of the fit
\begin{equation}\label{eq:poly_fit_copula}
    y = 0.1754 x^3 + 0.1119 x^2 - 0.1253 x + 0.1353.
\end{equation} Here $y$ represents $[\mathrm{Mg/Fe}]$ whereas $x$ is $[\mathrm{Fe/H}]$. This equation provides a simple way to split the two sequences in data space using copulas; we suggest using this split in other studies. The implications of this figure are discussed in detail in Section \ref{sec:results}. 

In the next section, we introduce the concept of elicitability and show how we can combine copulas and elicitable maps to estimate the correlation between age and metallicity as a continuous function of Galactocentric radius and distance from the midplane, $\rho(\tau, \mathrm{[Fe/H]} \mid R, z)$, of the Milky Way disk. We also use the two sequences split using the automated procedure described in this section to deduce the efficiency of secular evolution due to radial migration in the disk.
\begin{figure*}
    \centering
    \includegraphics[width=1.0\linewidth]{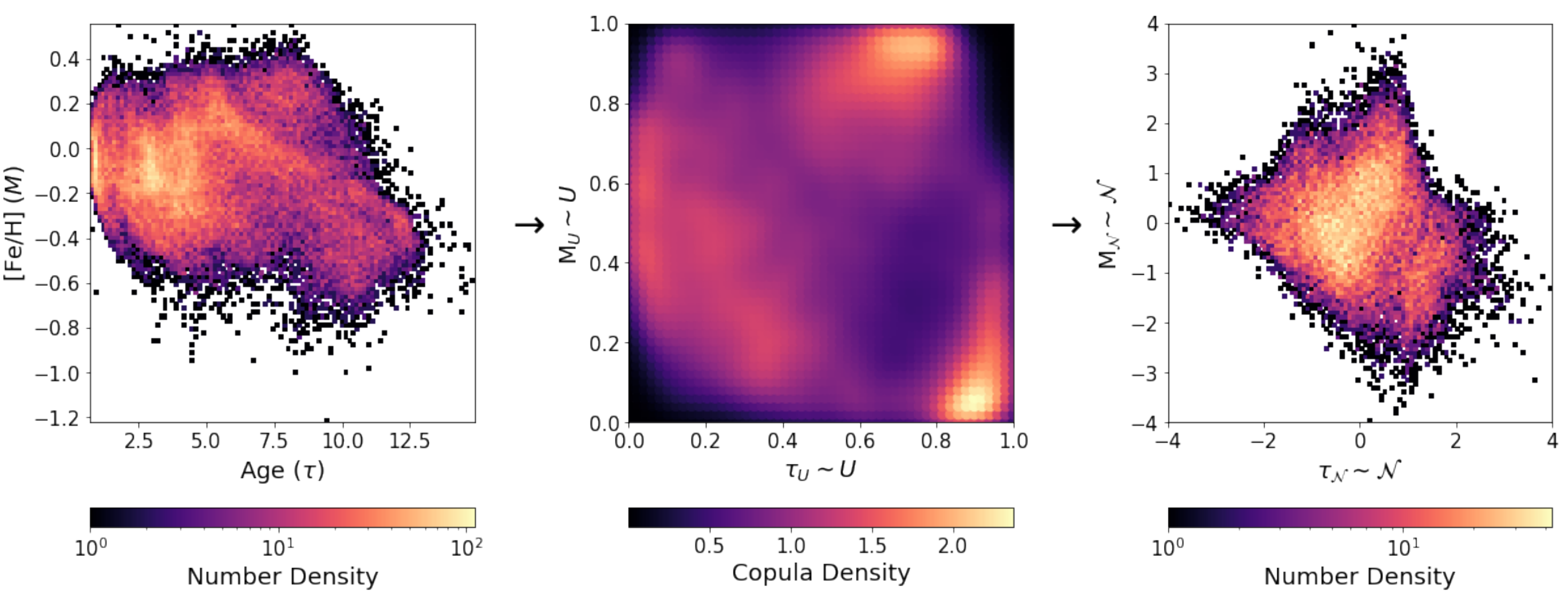}
    \caption{Transformation of the age-metallicity structure of our \emph{updated} stellar sample required to estimate the correlation between age and metallicity using a combination of elicitable maps and copulas. The left, middle, and right panels show the sample structure, its transformation to copula density, and the addition of Gaussian margins to the copula respectively.}
    \label{fig:gauss_marg_cop}
\end{figure*}

\section{Elicitable Maps}\label{sec:elicitable}
In this section, we study the age-metallicity relation across the Milky Way disk using copulas. To construct these copulas using the sparse data that we have from APOGEE, we introduce another novel statistical concept, that of elicitability. Elicitable maps allow the estimation of continuous functions of random variables using observed data instead of investigating binned properties. This allows us to obtain $\rho(\tau, [\mathrm{Fe/H}] \mid R, z)$ in the Milky Way disk.

Another advantage of these maps is the notion of using scoring functions to assess if point estimation is working well. To understand this, let us look at the example of estimating a statistical property $T$ of an underlying unknown distribution $F$ given some observations $X \sim F$. Instead of naively estimating $T(X)$, a corresponding score function allows us to rank the quality of an estimate. This is intimately tied to the idea of minimizing a loss function to train machine learning models. Thus, a good estimate or measure of dependence between stellar age and metallicity in the Galaxy should be the one that is least penalized by the relevant score function. 

Section \ref{subsec:math_elicit} describes the concept of elicitability and how it allows point estimation of a statistical function (or mapping) such as the mean. In Section \ref{subsec:joint_elicit}, we show that certain functions are not elicitable on their own (e.g., the variance), but are jointly elicitable with other mappings. Such higher order elicitability applies to the correlation function as well, and therefore this section acts as a basis for estimating the correlation between stellar age and metallicity $\rho(\tau, \mathrm{[Fe/H]})$ in this paper. Section \ref{subsec:cond_elicit} shows that we can also use elicitability to estimate conditional mappings, through which we estimate $\rho(\tau, [\mathrm{Fe/H}])$ conditional on $R$ and $z$ in the Galaxy or $\rho(\tau, [\mathrm{Fe/H}] \mid R, z)$ in Section \ref{subsec:cond_corr}. Finally, Section \ref{subsec:cond_corr_cop} combines copulas with conditional elicitability to obtain a more accurate measure of dependence between stellar age and metallicity compared with the standard (Pearson) correlation. Essentially, we show how to replace stellar ages and metallicities in $\rho(\tau, [\mathrm{Fe/H}] \mid R, z)$ with their copula space equivalents $U_\mathrm{age}, U_\mathrm{[\mathrm{Fe/H}]}$.

\subsection{Mathematical Background}\label{subsec:math_elicit}
An elicitable statistical map has the property that its point estimate may be obtained by minimizing a loss or \textit{scoring function} \citep{osband_1985, lambert_2008}. We describe this mathematically as follows.

Let $Y$ be a \textit{random variable} and $T: \R \to \R$ a statistical function or mapping of $Y$. Then, $T$ is an elicitable map if there exists a \textit{scoring function} $\S: \R \times \R \to \R$ that we can minimise to obtain the map $T$ as follows
\begin{equation}
     T(Y) = \underset{x \in \R}{\mathrm{arg\, min}} \;  \E [ \S(x, Y) ]\,.
\end{equation} 
To provide more intuition, let us look at the example of the mean. In this case, $T(Y) = \E [Y]$. Then, one score function $\S$ that elicits $T$ is
\begin{equation}\label{eq:mean}
    \S(x, Y) = (x - Y)^2\,,
\end{equation} 
i.e., the mean squared error. The following simple derivation illustrates how the mean squared error elicits the expectation
\begin{subequations}
\begin{align}
\E [ (x - Y)^2 ] & = x^2 - 2\,x\, \E [Y] + \E [Y^2]
\\
x^*=\underset{x \in \R}{\mathrm{arg\,min}} \; \E [ (x - Y)^2 ] &\Rightarrow \left.
\partial_x \,\E [ (x - Y)^2 ]\right|_{x=x^*} = 0
\\
&\Rightarrow 2\,x^* - 2 \,\E [Y] = 0
\\
&\Rightarrow x^* = \E [Y].
\end{align}
\end{subequations}
The above example demonstrates that with the right scoring function, we can perform point estimation of a desired elicitable statistical map. For more details, we refer the reader to \cite{gneiting_2011} upon which this section is based. We are interested, more generally, in eliciting the correlation between stellar age and metallicity conditional on $R$ and $z$ in the Galaxy $\rho(\tau, \mathrm{[Fe/H]} \mid R, z)$, for which we introduce the notion of \textit{higher-order elicitability} and its extension to \textit{conditional} mappings in the following two sections.

\subsection{Joint or Higher Order Elicitability}\label{subsec:joint_elicit}
A statistical map is elicitable if and only if it can be estimated by minimizing (the expectation of) a corresponding scoring function. As done for the mean above, one can prove that moments and quantiles are elicitable under mild assumptions \citep{savage_1971, thomson_1979, saerens_2000}. Such mappings are called $1$-elicitable since they can be estimated on their own. 

Several studies \cite[e.g.,][]{gneiting_2011, heinrich_2013} show that not all functions are 1-elicitable, a notable example is the variance. One can, however, jointly elicit the mean and variance, i.e., the pair (mean, variance) represents a $2$-elicitable mapping that is obtainable as the minimization of a score function that takes two inputs. This score function is given by
\begin{equation}\label{eq:mean_variance}
    \S(a,b; Y) = 
\frac{{a}^2 - 2\,b - 2\, a \, Y + Y^2}{b^2}
\end{equation} 
To demonstrate the above score estimates both mean and variance, we minimize 
\begin{equation}
    \E[\S(a,b ; Y)] = \frac{{a}^2 - 2\,b - 2\, a \, \E[Y] + \E[Y^2]}{b^2}
\end{equation}
over the inputs $a, b \in \R$. To do so, we equate the following first-order partial derivatives to zero. We then have
\begin{align}
    \partial_a \E[\S(a,b ; Y)] &= \frac{2}{b^2}\left(a-\E[Y]\right) = 0
    \\
    \partial_b \E[\S(a,b ; Y)] &= - \frac{2 (a^2 - b - 2 \,a\, \E[Y] + \E[Y^2])}{b^3} =0\,.
\end{align}
Solving the above equations gives the solution of the first-order conditions as
\begin{align}
    a = \E[Y] \qquad \text{and} \qquad b = \E[Y^2]-(\E[Y])^2.
\end{align} Thus, joint or higher order elicitability allows us to estimate several maps using a score that takes several inputs.

\cite{osband_1985} used the revelation principle to show that vector-valued mappings can be elicitable even though all their components are not $1$-elicitable \citep[also][]{gneiting_2011}. We can apply the revelation principle to the (mean, variance) because there exists a bijection between the pair and two 1-elicitable functions (the first two moments $\E [X]$ and $\E [X^2]$). In Section \ref{subsec:cond_corr}, we show that the correlation between two random variables is jointly elicitable with the means and variances of the two variables, thereby forming a 5-elicitable mapping corresponding to a score function with five inputs.

\subsection{Conditional Elicitable Maps}\label{subsec:cond_elicit}

The score function corresponding to an elicitable statistical mapping can also obtain the mapping conditional on a single, or collection of, other random variables \citep[refer to examples in][]{banerjee_2005}. For example, the score $\S$ corresponding to some statistical map $T$ can be used to obtain a point estimate of $T(Y \mid \bX)$, where $\bX$ is a $d$-dimensional random variable, by finding the minimizer
\begin{equation}\label{eq:cond_elicit}
    T [ Y \mid \bX ] = \underset{{g \in \mathcal{G} }}{\mathrm{arg\,min}} \; 
    \E \left[ \S( g(\bX), Y) 
    \right]
\end{equation} 
where the set $\mathcal{G}$ is the set of all square integrable functions from $\R^d \to \R$. That is, one seeks the function $g^*(\boldsymbol{x})$ that minimizes the score $\E [ \S( g(\bX), Y) ]$, and the function that minimises the score is the conditional map one seeks, i.e., $g^*(x)=T(Y|\bX=x)$. 

To further understand this, we again invoke the example of the mean $\E [Y]$ whose score function is the mean squared error. Consider that some information in $Y$ depends on other random variables $X$, and a function of $X$ quantifies this information. If we denote this function as $g(X)$, we find that the conditional expectation of $Y$ given $X$ is $\E [Z]$ \citep{banerjee_2005} or
\begin{equation}
    \E [ Y \mid X ] = \underset{{Z=g(X) :\, g \in \mathcal{G} }}{\mathrm{arg\,min}} \; \E [ ( Z - Y )^2]
\end{equation} The above equation is Equation \eqref{eq:cond_elicit} for the specific case of the statistical map being the conditional mean.

In practice, the expectation in Equation \eqref{eq:cond_elicit} is replaced by the empirical mean, i.e., one uses observed data to estimate the expected score and then seeks over functions to minimize it. More specifically, if we have observed data $(\bX^{(j)}, Y^{(j)})$ for $j = 1,...,N$, then we make the approximation
\begin{equation}
    \E [ \S(g(\bX), Y) ] \approx \frac{1}{N} \sum_{j=1}^N \S \left(g(\bX^{(j)}), Y^{(j)}\right).
\end{equation} 
The function $g$ that minimizes this sum is the estimate of the conditional map that we seek.

Seeking over all functions is not tractable in general, instead one seeks over a reduced set of functions. For example, in the case of $d=1$, we can use the set of $n$-degree polynomials: $g(x) = a_0 + a_1 x_1 + a_2 x^2 + \dotsc a_n x^n$ \citep{koenker_1978}. This approximation, however, may not work well in many cases, as the degree of the polynomial needs to be high, and then there is the potential of overfitting, and poor performance out-of-sample. Rather, to provide more flexibility in modeling $g$, we can parameterize it using an artificial neural network (NN) \citep[similar to][]{fissler_2023}. The loss function of the NN is the expected score that must be  minimized to estimate $g$, thereby obtaining an estimate of $T(Y \mid \bX)$ (more NN details in Section \ref{subsec:cond_corr_cop}).

The above shows that if a map $T(Y)$ is elicitable using a particular score $\S(x, Y)$, then the conditional version of that map $T(Y \mid \bX)$ is conditionally elicitable by replacing the argument $x$ with a function of a covariate $X$, so the new score is $\S( g(\bX), Y)$. We can also extend this approach to a set of maps that are jointly elicitable. The conditional versions of such maps are also elicitable, again, by replacing the arguments of the relevant score with functions. For example, for the (mean, variance) pair discussed in Section \ref{subsec:joint_elicit}, we estimate the conditional versions by rewriting the score in Equation \ref{eq:mean_variance} as $\S(g(\bX), h(\bX); Y)$.
 
Our next goal is to obtain an estimate for the jointly and conditionally elicitable $\rho(\tau, \mathrm{[Fe/H]} \mid R, z)$. The remaining element is to define the score function corresponding to this map.

\subsection{Conditional Correlation of Stellar Age-Metallicity}\label{subsec:cond_corr}
The Pearson's correlation coefficient quantifies the linear correlation between two random variables and is defined as the ratio between covariance and the product of standard deviations of the variables. The correlation between stellar age and metallicity can thus be written as 

\begin{equation}
\rho(\tau, \M) = \frac{\mathrm{cov}(\tau, \M)}{\sigma_{\tau}\, \sigma_{\M}}
\label{eq:rho}
\end{equation} Here $\M$ represents $[\mathrm{Fe/H}]$. We can express the above equation in expectations as

\begin{equation}
\rho(\tau, \M) = \frac{\E[\tau \, \M] - \E[\tau]\, \E[\M]}{\sqrt{\E[\tau^2] - (\E[\tau])^2\,} \;\sqrt{\E[\M^2] - (\E[\M])^2}\,}\,.
\label{eq:rho-from-means}
\end{equation} 
The above decomposition of the correlation is important because each of the components are moments and are elicitable even though the correlation alone is not -- rather it is only \textit{jointly} elicitable. 

When seeking the conditional correlation $\rho(\tau, \M \mid R, z)$, we modify all expectations in Equation \eqref{eq:rho-from-means} to their conditional ones, i.e., we make the replacements
\begin{subequations}
\begin{align}
        \E[\tau] &\rightarrow \E[\tau \mid R, z], & \E[\M] \rightarrow\; &\E[\M \mid R, z],
        \\
        \E[\tau^2] &\rightarrow \E[\tau^2 \mid R, z], &\E[\M^2] \rightarrow\; &\E[\M^2 \mid R, z], 
        \\
         \text{and} \quad \E[\tau \, \M] &\rightarrow \E[\tau \, \M \mid R, z].
\end{align}
\label{eq:cond_corr_comps}
\end{subequations}

Thus, to estimate the conditional correlation, we require estimates for the conditional mean and second moment of $\tau$ and $\M$ as well as the conditional cross-moment $\E[\tau\,\M \mid R, z]$, which can be obtained following the approach described in Section \ref{subsec:cond_elicit}. We can, instead, estimate the conditional means and second moments of $\tau$ and $\M$ and the conditional correlation directly. To achieve this goal, we use the score
\begin{equation}
\begin{split}
\S(a,b,c,d,f;\tau,\M) =\;& 
\frac{{a}^2 - 2\,b - 2 a \, \tau + \tau^2}{b^2} 
 \\
& + \frac{c^2 - 2 \,d - 2 c\, \M  + \M^2}{d^2} 
 \\
& + \left(f \,\sqrt{b\,d}+a\,c - \tau \, \M\right)^2\,,
\end{split}
\label{eq:score}
\end{equation}
and minimise the expected score (or loss)
\begin{equation}
\label{eq:loss}
\begin{split}
    \mathcal L := \E\big[
    \S\big(\mu_{\tau}(R,z), & \sigma^2_{\tau}(R,z),  \mu_{\M}(R,z), 
    \\
    &
    \sigma^2_{\M}(R,z),\rho_{\tau,\M}(R,z); \tau,\M \big)
    \big]    
\end{split}    
\end{equation}
over all functions $\mu_{\tau}$, $\sigma_{\tau}$, $\mu_{\M}$, $\sigma_{\M}$, $\rho_{\tau,\M}$ of two variables $(R,z)$ such that $\sigma_{\tau}$, $\sigma_{\M}$ are non-negative and $\rho_{\tau,\M}$ is in the interval $[0,1]$. We perform the minimization by parameterizing all of these functions with neural-networks and approximating the expectation with the empirical expectation. Thus, the individual neural networks estimate the means $\mu_{\tau}, \mu_{\M}$, variances $\sigma_{\tau}, \sigma_{\M}$, and correlation $\rho_{\tau,\M}$ conditional on $(R,z)$. Details of the architecture and mininimization procedure are deferred until the next section.

We explain how we obtain the score function in Equation \eqref{eq:score} by looking at the three terms individually. The first and second terms have the same form as Equation \eqref{eq:mean_variance} with $Y$ substituted by $\tau$ and $\M$ respectively, i.e., they together elicit $\mu_{\tau}, \sigma^2_{\tau}$ and $\mu_{\M}, \sigma^2_{\M}$ as indicated in Equation \eqref{eq:loss}. The third term is the same as Equation \eqref{eq:mean} (mean-squared error) with $Y$ and $x$ replaced by $\tau\,\M$ and $f \,\sqrt{b\,d}+a\,c$ respectively. Here, $f \,\sqrt{b\,d}+a\,c$ translates to the conditional expectation or cross-moment $\E[\tau\,\M \mid R, z]$ when using the functions in Equation \eqref{eq:loss} and its parameterization using $f$ allows us to directly estimate the conditional correlation. Overall, this means that the conditional means and variances of $\tau$ and $M$ and the conditional correlation between the two quantities are 5-elicitable and can be estimated using the score function in Equation \eqref{eq:score}. 

Note that the first two terms in Equation \eqref{eq:score} are independent of each other and can be minimized separately to obtain $\mu_{\tau}$, $\sigma_{\tau}$ and $\mu_{\M}$, $\sigma_{\M}$ respectively. The third term, however, ties all three together, and minimizing over all five quantities using a single loss function leads to a consistent estimation of all quantities.

This procedure yields an estimate of the spatial distribution of stellar age and metallicity correlation in the Galaxy, i.e., $\rho(\tau, \M \mid R, z)$. In the next section, we incorporate copulas into this estimation procedure to obtain a measure of dependence between stellar age and metallicity that is not distorted by the marginal distribution of $\tau$ and $\M$.

\subsection{Conditional Correlation of Age-Metallicity using Copulas}\label{subsec:cond_corr_cop}

The results of section \ref{subsec:math_cop} demonstrate that copulas provide scale-invariant and non-linear measures of dependence between random variables as opposed to the scale-dependent correlation. Thus, instead of estimating the correlation $\rho(\tau, \M \mid R, z)$, we estimate the correlation of the transformed variables in copula space -- this results in what is called the Spearman's correlation or rank correlation. To do so, we adopt the following methodology (as illustrated in Figure \ref{fig:gauss_marg_cop}):
\begin{enumerate}
    \item From the \emph{updated} stellar samples of age and metallicity  $(\tau^{(j)}, \M^{(j)})_{j\in \J}$, where $\J=\{1,\dots,J=22,527\}$, obtain the empirical CDFs $F_\tau(\cdot)$ and $F_M(\cdot)$ \label{step:1}
    
    \item Transform the data into copula space $\tau^{(j)}, \M^{(j)} \rightarrow \tau_U^{(j)}=F_\tau(\tau^{(j)}), \M_U^{(j)}=F_M(M^{(j)})$ \label{step:2}
    
    \item Transform the uniformly-distributed data $\tau_U^{(j)}$ and $\M_U^{(j)}$ from (ii) into Gaussian margins to obtain $\tau_\N^{(j)}, \M_\N^{(j)}$ \label{step:3}
    
    \item Estimate $\rho(\tau_\N, \M_\N \mid R, z)$ using $\{\tau_\N^{(j)}, \M_\N^{(j)}, R^{(j)}, z^{(j)}\}$. \label{step:4}
\end{enumerate}
Steps \ref{step:1} and \ref{step:2} is the procedure we use to generate the copula in the previous section. Step \ref{step:3} transforms the data into data with Gaussian marginals, but retains the same copula as the original data. 
\tikzstyle{circle}=[shape=circle,draw=green,fill=blue!10]
\tikzstyle{box_lime}=[shape=rectangle,draw=black!50,fill=lime!20]
\tikzstyle{box_blue}=[shape=rectangle,draw=black!50,fill=blue!20]
\tikzstyle{box_red}=[shape=rectangle,draw=black!50,fill=red!20]
\tikzstyle{box_empty}=[shape=rectangle,draw=black!50]

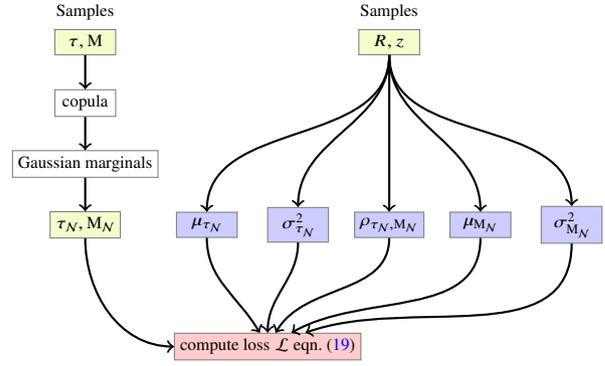
\begin{figure}
    \centering
    \begin{tikzpicture}[scale=0.8,every node/.style={transform shape},minimum width=1.0cm]

    \node at (-3,10.5) {Samples};
    \node[box_lime] (samples) at (-3,10) {$\tau,\M$};

    \node[box_empty] (copula) at (-3,9) {copula};
    \draw [thick,->] (samples) to (copula);

    \node[box_empty] (gaussian) at (-3,8) {Gaussian marginals};
    \draw [thick,->] (copula) to (gaussian);

    \node[box_lime] (samples-N) at (-3,7) {$\tau_\N,\M_\N$};
    \draw [thick,->] (gaussian) to (samples-N);

    \node at (2,10.5) {Samples};
    \node[box_lime] (samples-rz) at (2,10) {$R,z$};

    \node[box_blue] (NN1) at (-1,7) {$\mu_{\tau_\N}$};
    \node[box_blue] (NN2) at (0.5,7) {$\sigma^2_{\tau_\N}$};
    \node[box_blue] (NN3) at (2,7) {$\rho_{\tau_\N,\M_\N}$};
    \node[box_blue] (NN4) at (3.5,7) {$\mu_{\M_\N}$};
    \node[box_blue] (NN5) at (5,7) {$\sigma^2_{\M_\N}$};    

    \draw [thick,->] (samples-rz) to [out=-90,in=90] (NN1) [];
    \draw [thick,->] (samples-rz) to [out=-90,in=90] (NN2) [];
    \draw [thick,->] (samples-rz) to (NN3) [];
    \draw [thick,->] (samples-rz) to [out=-90,in=90] (NN4) [];
    \draw [thick,->] (samples-rz) to [out=-90,in=90] (NN5) [];

    \node[box_red] (loss) at (0,5) {compute loss $\mathcal{L}$ eqn. \eqref{eq:loss}};

    \draw [thick,->] (NN1) to [out=-90,in=120] (loss) [];
    \draw [thick,->] (NN2) to [out=-90,in=90] (loss) [];
    \draw [thick,->] (NN3) to [out=-90,in=60] (loss) [];
    \draw [thick,->] (NN4) to [out=-90,in=30] (loss) [];
    \draw [thick,->] (NN5) to [out=-90,in=20] (loss) [];    
    \draw [thick,->] (samples-N) to [out=-90, in=180] (loss) [];
    
    \end{tikzpicture}
        \caption{The general structure of the procedure for estimating the correlation of age $\tau$ and metallicity $\M$ conditional on Galactocentric radius $R$ and absolute distance from midplane $|z|$ using our APOGEE \emph{updated} stellar sample. The blue rectangles represent NNs that accept $R,z$ as features. The loss in the red rectangle is minimized by grabbing minibatches of data (nodes that feed into the loss), using backpropogation to compute gradients w.r.t. NN parameters, and taking AdamW update steps.}
    \label{fig:nn_arhcitecture}
    
\end{figure}

To estimate the conditional correlation, we require five neural nets that feed into the loss function in \eqref{eq:loss}. These NNs correspond to estimates of
\begin{enumerate}
    \item $\E[\tau_\N \mid R, z]$
    \item $\E[\M_\N \mid R, z]$
    \item $\E[\tau_\N
^2 \mid R, z] - (\E[\tau_\N
 \mid R, z])^2$
    \item $\E[\M_\N
^2 \mid R, z] -(\E[\M_\N
\mid R, z])^2$
    \item $\rho(\tau_\N
, \M_\N \mid R, z)$.
\end{enumerate} 
and we denote the estimates by $\mu_{\tau_\N}$, $\mu_{\M_\N}$, $\sigma_{\tau_\N}^2$, $\sigma_{\M_\N}^2$, and $\rho_{\tau_\N, \M_\N}$ respectively.  
For the NN archictecure, we use feedforward NNs with 5 hidden layers, 20 nodes each, and apply sigmoid linear unit (SiLU) activation functions in all internal layers. For the output layer, the NNs corresponding to means have no activation (as the mean can take any value in $\R$), the ones corresponding to variances have softplus output activation functions (as variances must be non-negative), and the one corresponding to correlation has a $\tanh$ output activation (as correlations must lie in $[-1,1]$).

As discussed in Section \ref{subsec:cond_corr}, the various conditional quantities are estimated by seeking over all functions (in our case the functions are approximated by the NN architectures mentioned above) to minimize the loss function \eqref{eq:loss}. We do so by grabbing minibatches of $256$ from the total of $N=22,527$ data samples of $(\tau_\N, \M_\N, R, z)$, pushing the $R,z$ pairs through the NNs, and then estimating the loss \eqref{eq:loss} using the empirical mean. 

\begin{figure}
    \includegraphics[width=1.0\linewidth]{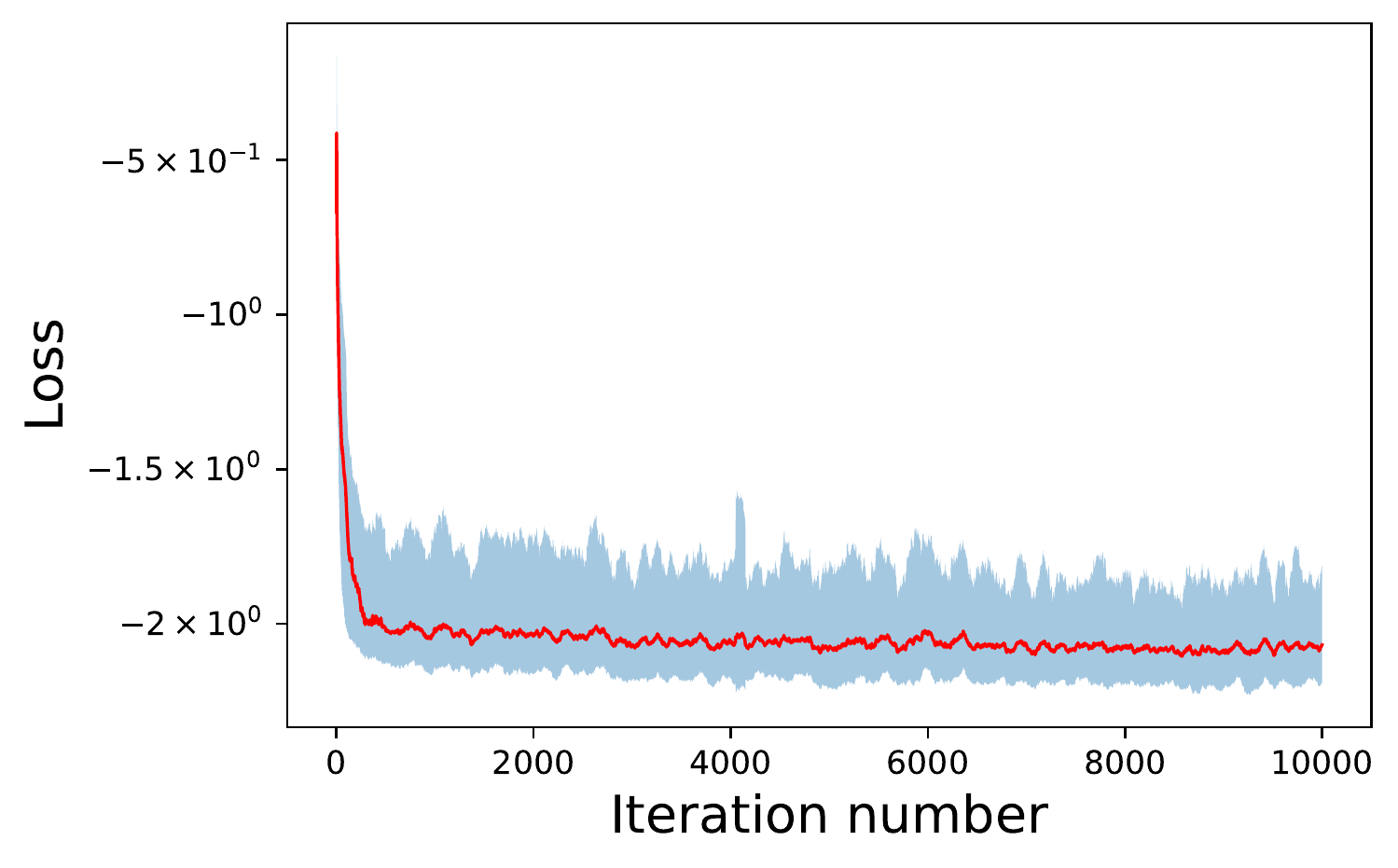}
    \caption{The moving average of the loss of the conditional correlation neural network model (Figure \ref{fig:nn_arhcitecture}) along with its uncertainty using the moving standard deviation.}
    \label{fig:loss}
\end{figure}

We minimize the loss (or the expected score) by using the \texttt{AdamW} \citep{loshchilov_2018} optimizer with a learning rate of 0.005 using the \texttt{PyTorch} module \citep{pytorch}. \texttt{AdamW} is an improved version of Adam \citep{kingma_2015} that requires backward propagating the loss \citep{loshchilov_2018}. Additionally, we use a learning rate scheduler called \texttt{StepLR} that adjusts the rate every few epochs and helps stabilize the optimization results. Figure \ref{fig:nn_arhcitecture} illustrates this training procedure. We monitor the learning rate hyper-parameter along with the loss and find that the loss converges after about $1,000$ iterations (refer to Figure \ref{fig:loss}).

\begin{figure*}
    \centering
    \includegraphics[width=0.6\linewidth]{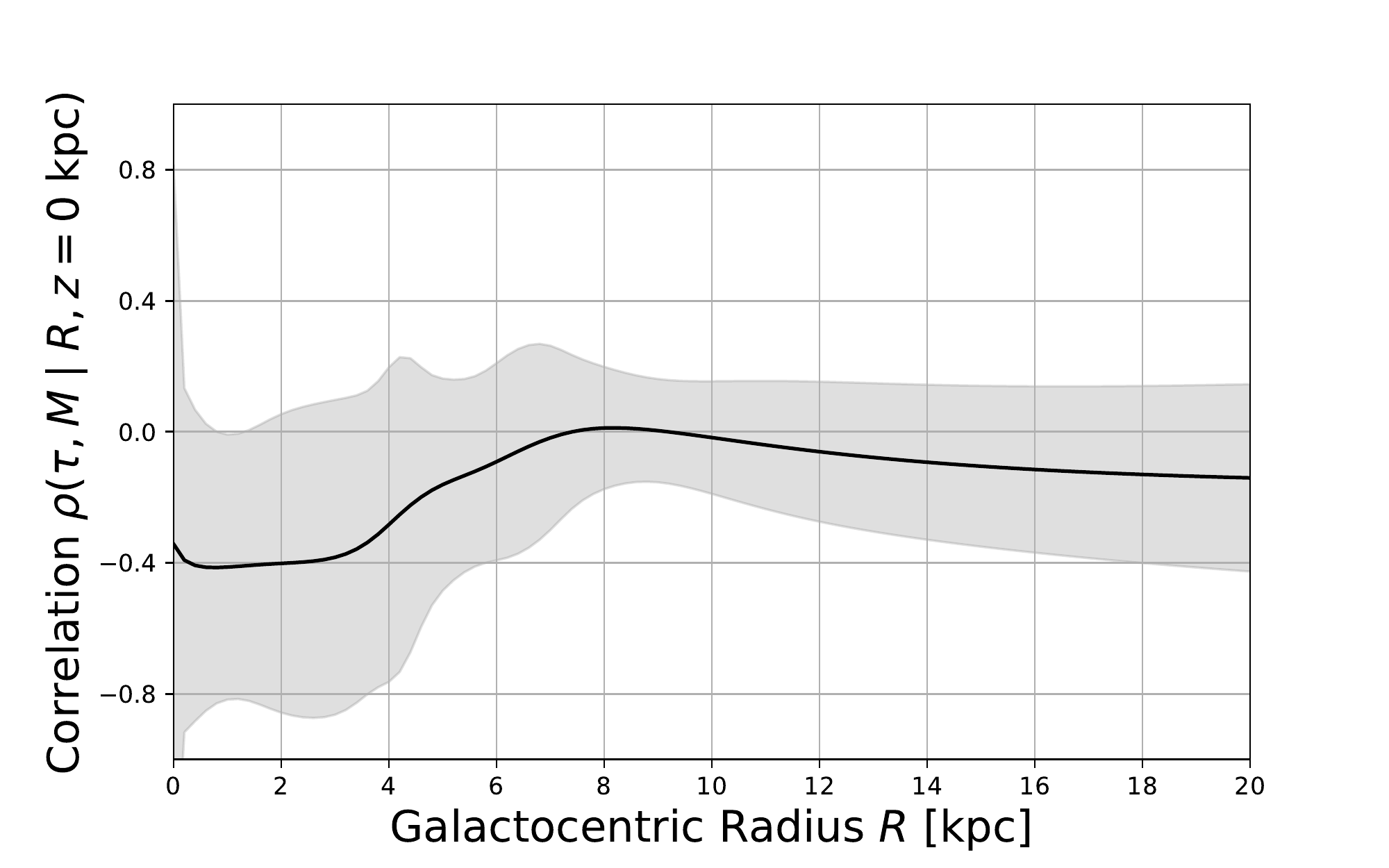}
    \caption{The correlation between stellar age and metallicity as a function of Galactocentric radius $R$ at the midplane ($z=0$ kpc) of the disk derived using copulas and elicitable maps. The grey shaded region shows the uncertainties on the correlation estimate, particularly, the 68\% confidence intervals using 1 $\sigma_J$ (jackknife variance).
    }
    \label{fig:age-metallicity-z=0}
\end{figure*}

\begin{figure*}
    \includegraphics[width=1.0\linewidth]{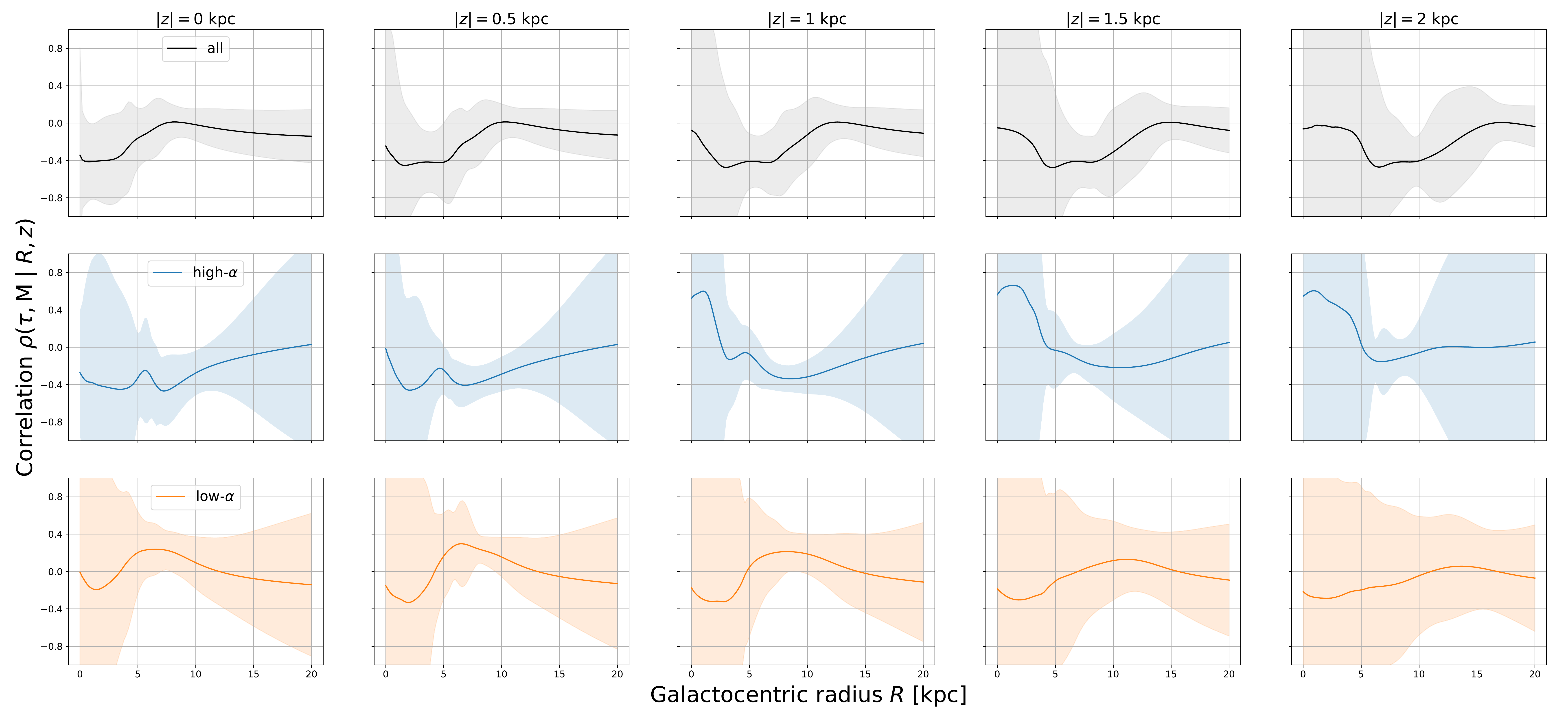}
    \caption{The correlation between stellar age and metallicity as a function of Galactocentric radius $R$ at different heights $|z|$ above and below the midplane of the disk. The top panels show the correlation function for the \emph{updated} sample whereas the middle and bottom panels display that for the high-$\alpha$ and low-$\alpha$ sequences (using the split in Section \ref{subsec:septracks}) respectively.}
    \label{fig:age-metallicity-z=all}
\end{figure*}

Figure \ref{fig:age-metallicity-z=0} shows our learned correlation function $\hat{\rho}$ sampled at the midplane of the disk, i.e., $z=0$. In addition to $\hat{\rho}$, we provide uncertainties on the estimated function by estimating its variance using jackknife resampling \citep{quenouille_1949}. Essentially, we apply the following procedure:
\begin{enumerate}
    \item Subdivide $\S = \{\tau_\N^{(j)}, \M_\N^{(j)}, R^{(j)}, z^{(j)} \mid j=1,\dotsc,\;22,527\}$ into $K \sim 50$ equal-sized subsets $\S_1, \dotsc, \S_K$
    
    \item Remove or delete each subset $\S_i$ from the whole set $\S$ to generate $K$ groups
    \begin{equation}
        \S_{\setminus i} = \{\S_1 \cup \dotsc S_{i-1} \cup \S_{i+1} \cup \dotsc \S_K\} \; \; \; \mathrm{for} \; i =1, 2,\dotsc K
    \end{equation}
    \item Train $\rho(\tau_\N, \M_\N \mid R, z)$ on each group $\S_{\setminus i}$ to obtain the delete-one estimates $\Hat{\rho}_{\setminus i}$    
    \item Finally, estimate the jackknife variance $\sigma^2_J$ of $\hat{\rho}$ as
    \begin{align}\label{eq:jk_var}
        \sigma^2_J(\Hat{\rho}) &= \frac{K-1}{K} \sum_{i=1}^K (\Hat{\rho}_{\setminus i} - \Hat{\rho}_{\setminus \bullet})^2 \; \; \; \mathrm{where}\\
        \Hat{\rho}_{\setminus \bullet} &= \frac{1}{K} \sum_{i=1}^{K} \Hat{\rho}_{\setminus i}
    \end{align} and the confidence interval using the standard deviation $\sigma_J$ as
    \begin{equation}
        \mathrm{CI} = \left(\Hat{\rho}_{\setminus\bullet} - \sigma_J(\Hat{\rho})
        \;;\;
        \Hat{\rho}_{\setminus\bullet} + \sigma_J(\Hat{\rho}) \right)
    \end{equation}
\end{enumerate} Thus, $\sigma_J(\Hat{\rho})$ provides a measure of uncertainty on the correlation estimate. We plot the above confidence interval in Figure \ref{fig:age-metallicity-z=0}.

Note that we use $K \sim 50$ jackknife subsets because the variance $\sigma^2_J$ in Equation \eqref{eq:jk_var} remains stable beyond this $K$. While jackknifing should provide a conservative estimate of the variance, we note that it can be significantly inflated in the presence of outliers \citep{miller_1974}. Thus, the reported uncertainties could potentially be larger than those underlying the true distribution. In our data, we do find such outliers, hence, suspect that the estimated confidence bands are indeed larger than the true confidence bands.

In addition to $z=0$, we investigate the age-metallicity correlation function at different distances above and below the midplane of the disk, and found that this relation is symmetric about $z=0$. We thus updated our NNs to estimate $\rho(\tau_\N, \M_\N \mid R, |z|)$, i.e., the age-metallicity correlation conditional on $R$ and the absolute value $|z|$. The top panels of Figure \ref{fig:age-metallicity-z=all} display this correlation function at different heights $|z| \in [0, 2]$ kpc.

We also study the age-metallicity correlation of the high- and low-$\alpha$ sequences separately using the split obtained in Section \ref{subsec:septracks}. Essentially, we apply Steps \ref{step:1}, \ref{step:2}, and \ref{step:3} to transform the two sequences, $\tau_\mathrm{high}^{(j)}, \mathrm{\M}_\mathrm{high}^{(j)}$ for $j=1,\dotsc,4201$ and $\tau_\mathrm{low}^{(j)}, \mathrm{\M}_\mathrm{low}^{(j)}$ for $j=1,\dotsc,18326$, as shown in Figure \ref{fig:gauss_marg_cop}. We then use the transformed data $\N_{\tau, \mathrm{high}}, \N_{\M, \mathrm{high}}$ and $\N_{\tau, \mathrm{low}}, \N_{\M, \mathrm{low}}$ to learn the spatial distribution of age-metallicity correlation for the two sequences. The middle and bottom panels of Figure \ref{fig:age-metallicity-z=all} display the learned correlation for the high- and low-$\alpha$ sequences separately. We discuss these results in detail in the next section.

 \section{Results \& Discussion}\label{sec:results}
Section \ref{subsec:septracks} presents an automated copula-based approach to disentangling the dependency between stellar abundances $[\mathrm{Fe/H}]$ and $[\alpha\mathrm{/Fe}]$, and determining the dichotomy in the high- and low-$\alpha$ disk sequences. Section \ref{subsec:alpha_discuss} compares the sequences we obtain with those in the literature and discusses its implications. Section \ref{subsec:cond_corr_cop} demonstrates the applicability of elicitable maps to estimate the correlation between stellar age and metallicity as a continuous function of $R, z$ in the disk. Particularly, we evaluate this correlation separately for our estimated $\alpha$ sequences. In Section \ref{subsec:age-metallicity_discuss}, we use our correlation results to empirically understand the efficiency of radial migration in different parts of the disk.

\subsection{High- and Low-$\alpha$ Disk}\label{subsec:alpha_discuss} 
\begin{figure}
    \centering
    \includegraphics[width=\linewidth]{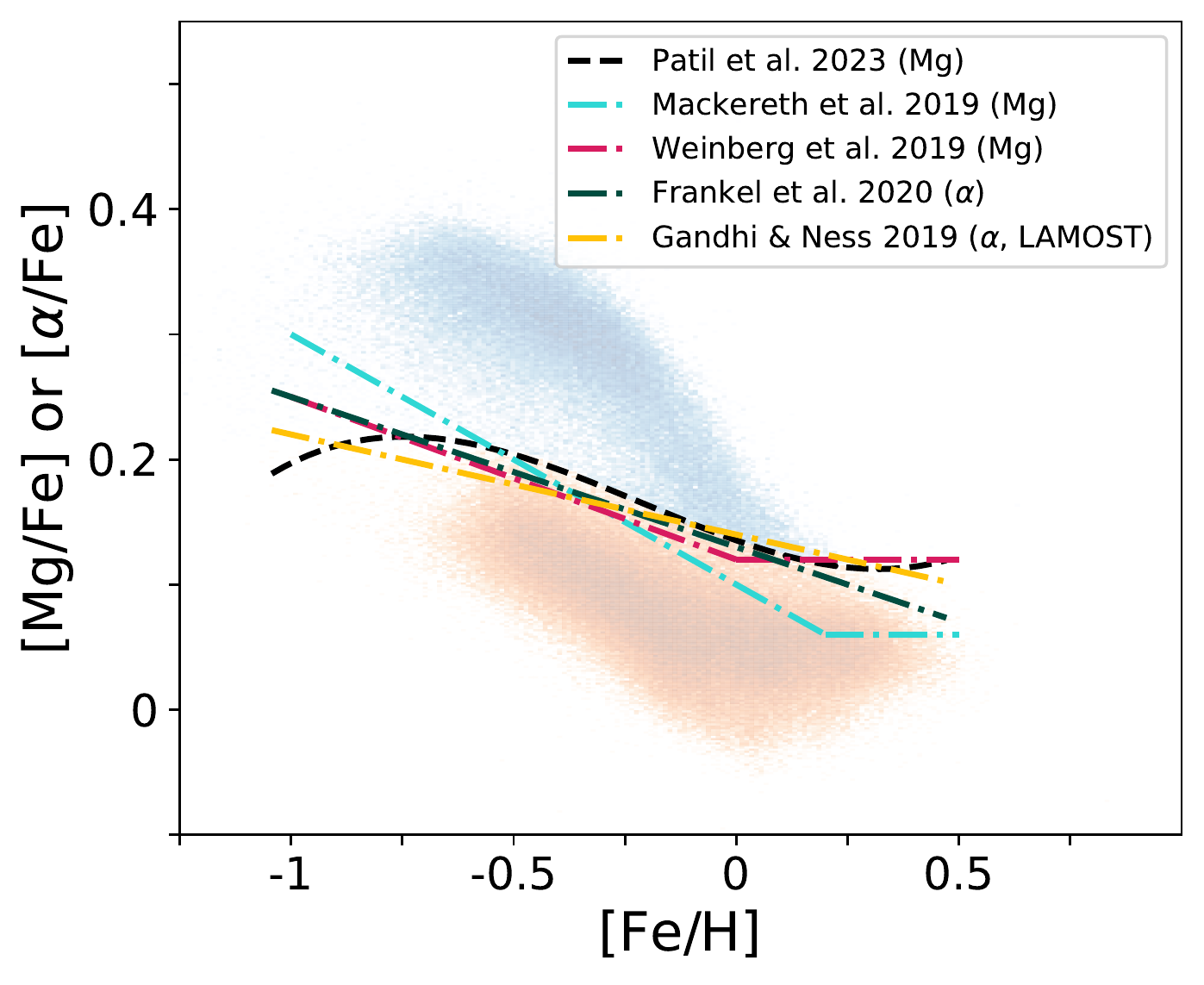}
    \caption{Comparison of our split of $\alpha$ sequences (black dashed line) with recent ones in the literature (dash-dotted lines). Three example splits shown in turquoise and magenta \citep{mackereth_2019_accretion, weinberg_2019, frankel_2020} are based on APOGEE data, whereas the one in yellow \citep{gandhi_2019} uses the LAMOST survey. Note that \citealt{gandhi_2019, frankel_2020} directly use $\alpha$ as opposed to the others that use magnesium as a proxy for $\alpha$ (similar to this paper).}
    \label{fig:splits_compare}
\end{figure}

In Figure \ref{fig:MgFe_copula}, we clearly observe the bimodality of the Milky Way disk in $[\alpha\mathrm{/Fe}]-[\mathrm{Fe/H}]$ abundance space, particularly at low metallicity. The two $\alpha$ sequences intersect at super-solar metallicity. However, it is unclear whether the metal-rich intermediate-age stars belong to the high-$\alpha$, low-$\alpha$, or an overlap between the two disks. Nevertheless, most studies that investigate the evolutionary history of the two-component disk manually divide the $\alpha$ sequences by drawing a line or polynomial in the $([\alpha\mathrm{/Fe}], [\mathrm{Fe/H}])$ space \citep[e.g.,][]{nidever_2014, weinberg_2019, bland-hawthorn_2019, gandhi_2019, mackereth_2019_accretion, ness_2019, frankel_2020, leung_2023} based on visual inspection. Some studies include dynamical cuts that help separate the two-component disk \citep[e.g.,][]{xiang_2022}, but these cuts are still based on manual selection. This division is subjective, often dependent on sample selection and stellar parameter estimates, and is therefore inconsistent across studies. We illustrate that manual separations lead to discrepancies by plotting a few examples in Figure \ref{fig:splits_compare}. This figure also shows our split which we obtain in an automated fashion by looking at the copula of the stellar ($[\alpha\mathrm{/Fe}], [\mathrm{Fe/H}]$) plane. Our split seems to be the most consistent with \cite{gandhi_2019}, whose resulting sequences are shown to be dynamically distinct in orbital actions $J_R, J_z,$ and $J_\phi (L_z)$. This illustrates that our purely data-driven approach has physical backing and is not necessarily specific to a certain survey (after accounting for differences in calibration).

\begin{figure}
    \includegraphics[width=1\linewidth]{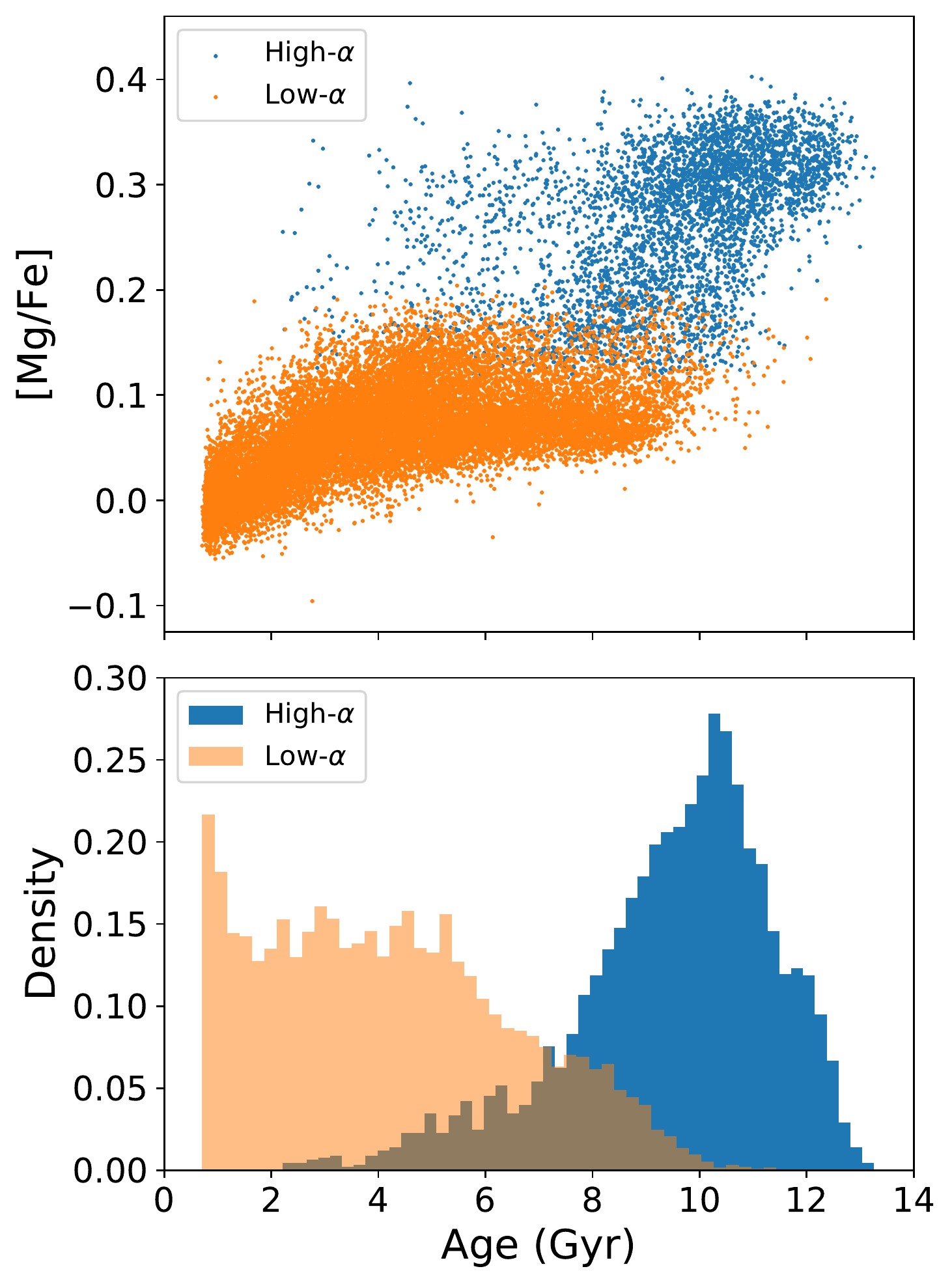}
    \caption{$[\mathrm{Mg/Fe}]$ versus age distribution (top panel) and age histograms (bottom panel) of the low- and high-$\alpha$ sequences we obtain by applying our copula split (Figure \ref{fig:split_data}) to the \emph{updated} sample. We see that the age dependence of the two sequences is relatively distinct, suggesting a sequential formation scenario.}
    \label{fig:split_data_ages}
\end{figure}

In order to understand the dependence in age of the two-component disk, we apply our split to the \emph{updated} sample with precise ages. The resulting high- and low-$\alpha$ sequences have $4201$ and $18326$ stars respectively. Their age distributions are shown in the bottom panel of Figure \ref{fig:split_data_ages}. The distributions are fairly distinct with the high-$\alpha$ sequence concentrated in the age range of ${\sim}7.5 - 12.5$ Gyr, and the low-$\alpha$ between $0-7.5$ Gyr. There is some overlap between $7.5 - 10$ Gyr which could be alleviated when considering typical age uncertainties. The youngest stars in the high-$\alpha$ sequence have similar ages as the oldest ones in low-$\alpha$. Note that our approach includes the (super-)metal-rich, intermediate age stars in the low-$\alpha$ sequence \citep[in contrast with the suggestion in][]{feuillet_2019}, which potentially removes some of the difficulties in explaining the $[\alpha\mathrm{/Fe}]$ evolution of the Galactic disk. Rapid star formation is required to significantly increase $\alpha$ through Type II supernovae in comparison with $\mathrm{Fe}$ \citep{chiappini_2001}. This process is harder in later stages of disk evolution when Type Ia supernovae are abundant. If the metal-rich low-$[\alpha\mathrm{/Fe}]$ stars are part of the high-$\alpha$ sequence, then the low-$\alpha$ disk needs to account for a large increase in $[\alpha\mathrm{/Fe}]$ as $[\mathrm{Fe/H}]$ reduces. However, this is not the case in our split of low-$\alpha$ disk stars and therefore its evolution can be explained by more gradual star formation where the two sequences essentially form sequentially in time. Note that previous studies show that the kinematics of the metal-rich, intermediate $\alpha$ (or age) stars are consistent with those of the geometric thin disk \citep{hayden_2017}, and these results are in line with our results based on the ($[\alpha\mathrm{/Fe}], [\mathrm{Fe/H}]$) copula space.

Our method is the first automated separation of the high- and low-$\alpha$ disk using copulas. However, there have been previous data-driven approaches for investigating this chemical bimodality in the disk. For example, \cite{anders_2018} use t-distributed stochastic neighbour embedding (t-SNE) on a multi-dimensional abundance space and show that the above discussed metal-rich stars represent sub-populations that are closer to the thin disk stars in t-SNE space. These sub-populations, called Inner Disk III and IV, are expected to have radially migrated from the inner parts of the disk and are now on cold orbits \citep{kordopatis_2015}. Therefore, their abundance and kinematic properties are similar to low-$\alpha$ disk stars, which is consistent with our split of the low-$\alpha$ sequence. Thus, even with the inclusion of more abundances, their results have some similarities with our split. There have also been other data-mining approaches that reduce the dimensionality of multi-abundance spaces \citep{ting_2012, patil_2022a, ratcliffe_2020}. In the future, we would like to extend our copula approach to a higher-dimensional abundance space to asses whether that further improves the separation of the $\alpha$ sequences. We would also like to include kinematic information such as the azimuthal action $J_\phi$ (or angular momentum $L_z$) along with $[\alpha\mathrm{/Fe}]$, $[\mathrm{Fe/H}]$ to further explore the $\alpha$ dichotomy \citep[as done in e.g.,][]{xiang_2022}.

Note that we automate the splitting of $\alpha$ sequences using contours in copula space because this space shows a cleaner separation. The same is not the case in abundance space where the two sequences merge. Therefore, applying clustering algorithms directly to abundances and automatically splitting the sequences is challenging. For example, \cite{ratcliffe_2020} apply a hierarchical clustering method to find populations in a 19 abundance space. They project their two resulting clusters onto the $[\alpha\mathrm{/Fe}]-[\mathrm{Fe/H}]$ plane and show that some of the metal-rich stars that are generally included in the high-$\alpha$ sequence are more grouped with the low-$\alpha$ sequence. However, when they increase the number of clusters, these stars seem to be a separate population, and it is unclear how many clusters are truly present even with the inclusion of multiple abundances. Another study that investigates splitting the $\alpha$ sequences using clustering is \cite{blancato_2019}. They fit a Gaussian mixture model to the $[\alpha\mathrm{/Fe}]-[\mathrm{Fe/H}]$ plane to extract two clusters. They find that the clustering is sensitive to model initialization, and end up having to manually pick the approximate center locations of the two sequences. Thus, simple unsupervised clustering methods in abundance space do not work as well as our copula based approach. 

It is important to mention that clustering could also be applied to split the sequences in copula space instead of our contour-based approach. We perform some tests using different clustering algorithms (e.g., k-means, DBSCAN, Gaussian mixture) and find that the non-linearity of the separation makes it hard for them to locate the sequences. In the future, we would like to compare our approach with more involved clustering algorithms and other gap/valley finding methods \cite[e.g.,][ and the flow line approach discussed in Section \ref{subsec:septracks}]{contardo_2022}.

We further explain the implications of our split in the next section, which demonstrates that the age-metallicity structures of the resulting $\alpha$ sequences show distinct features.
\begin{figure*}
    \includegraphics[width=1.0\linewidth]{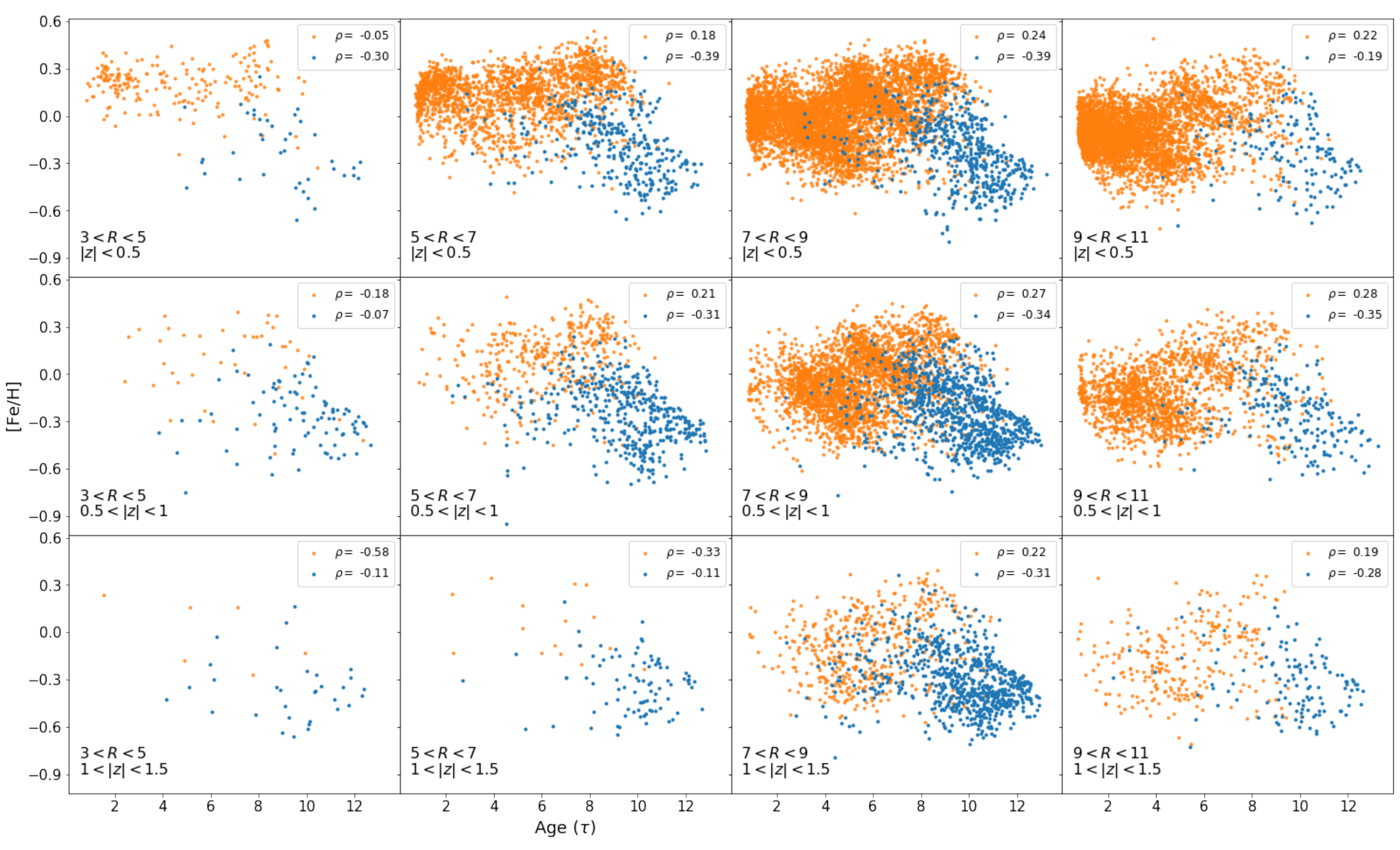}
    \caption{Distributions of stars in the age-metallicity plane for the high- and low-$\alpha$ sequences (in blue and orange respectively) as a function of position, shown using 12 disk zones binned in $R$ and $|z|$. Age-metallicity correlation $\rho$ estimates for the sequences in each disk zone are also shown.}
    \label{fig:age-metallicity-naive}
\end{figure*}
\subsection{Spatial distribution of the age-metallicity relation}\label{subsec:age-metallicity_discuss}
Studies have repeatedly shown that stars of the same age in the solar neighbourhood have a wide range of metallicities \citep{edvardsson_1993, haywood_2006, casagrande_2011, bensby_2014} instead of the tight relation between age and metallicity expected by one-zone models of chemical evolution. \cite{feuillet_2019} observe this trend across 12 disk zones (e.g. $7 < R < 9$ kpc and $|z| <0.5$ kpc) by estimating mean stellar ages in different metallicity bins. The fact that there is large variance in $p([\mathrm{Fe/H}])$ at any given age means that age and metallicity are not strongly correlated with each other, i.e. when $[\mathrm{Fe/H}]$ increases, age does not necessarily decrease. This lack of strong correlation (combined with the spatial metallicity gradient in the disk) is a potential signature of stellar radial migration. Here we present the first investigation of the spatial distribution of correlation between stellar age and metallicity as a continuous function of $R$ and $z$. This provides a fine-grained view of the variations in the age-metallicity relation across the disk instead of looking at trends in binned disk zones.

Figure \ref{fig:age-metallicity-z=0} shows the age-metallicity correlation across Galactocentric radius at the midplane of the disk $\rho(\tau, \M | R, |z|=0)$ in black  (we drop the subscript $\tau_\N, \M_\N$ for simplicity). The light grey region provides a measure of uncertainty through jackknife sampling. We see that the correlation is consistent with being zero from $R = 5$ to $20$ kpc suggesting that radial migration is efficient near the plane of the disk, particularly near the Sun and the outer parts of the disk. We also observe a decrease in the age-metallicity correlation inwards of the solar neighbourhood, which settles to a constant value of $-0.4$ at $R \lesssim 5$ kpc. 

The above result is consistent with theoretical predictions and observations \citep[e.g.,][]{sellwoodbinney2002, schonrich_binney_2009a, frankel_2018, frankel_2020}; these studies show that radial migration preferentially affects stars on near-circular kinematically cold orbits, which generally describe the motion of young stars in the plane of the disk. Such orbits undergo large variations in angular momentum $L_z$ without corresponding increase in random energy (or eccentricity/vertical extent) when they are in resonance with non-axisymmetric forces (e.g., spiral perturbations and bar) in the disk. Thus, they move or migrate significantly from their birth radii as opposed to warm eccentric orbits that experience coupled variations in $L_z$ and random motion. Young disk stars on nearly circular orbits are predominantly near the sun and the outer parts of the disk. Their radial migration inwards or outwards in the disk where the gas exhibits higher or lower enrichment levels (negative radial metallicity gradient) dilutes the stellar age-metallicity correlation in these disk regions.

\cite{minchev_2010} and \cite{minchev_2011} show that more efficient radial migration is possible when there is overlap between bar and transient spiral resonances, which leads to further reduction of the age-metallicity correlation. In addition, there are fewer stars in the outer parts of the disk as compared to the centre, and therefore more effects of radial migration on this correlation should be seen at large $R$. We observe these trends in Figure \ref{fig:age-metallicity-z=0}. 

\begin{figure*}
    \includegraphics[width=\linewidth]{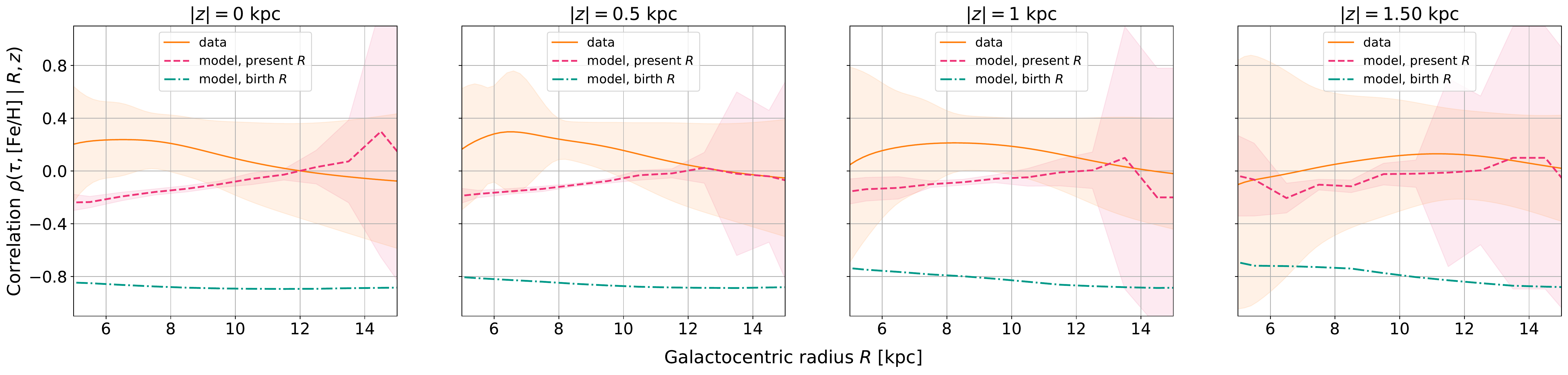}
    \caption{Comparison between the stellar age-metallicity correlation estimated using APOGEE data (orange) in this study and that using the mock sample from a radial migration model \citep{frankel_2020}. The green dash-dotted line shows this correlation when stars are at their birth radii in the model, i.e., without the effects of radial migration. The magenta dashed line shows the correlation at present radii, i.e., with radial migration. We see that our results are consistent with those from the radial migration model, suggesting that this evolutionary process is significant in the Galactic disk.}
    \label{fig:frankel_compare}
\end{figure*}

At $R \lesssim 5$ kpc, we observe a negative correlation between stellar age and metallicity in Figure \ref{fig:age-metallicity-z=0}. This central region of the Galaxy corresponds to the location of the bar, where orbits of stars are far from circular \citep{bovy_2019}. We do not expect radial migration to occur inside the bar, which explains the tighter correlation we observe in our results.

We now apply the above reasoning to understand the variations in the stellar age-metallicity correlation with increasing distance $|z|$ from the midplane. In the top panels of Figure \ref{fig:age-metallicity-z=all}, we plot the learned $\rho(\tau, \M \mid R, |z|) \; \mathrm{for} \; |z| \in \{0, 0.5, 1, 1.5, 2\}$ kpc. We see that the age-metallicity correlation is still consistent with being $-0.4$ in the inner disk, but the radius $R$ at which the correlation becomes zero potentially increases with vertical distance $|z|$. One interpretation of this result is that at low $R$, stars at a given $z$ are less susceptible to radial migration than at larger $R$. Another possible interpretation is that at low $R$, the age-metallicity relation does not vary much with radius, so it is conserved under radial migration; studies have shown that the observed $[\mathrm{Fe/H}]$(R) profile is flatter in the inner disk \citep[e.g.,][]{pietrukowicz_2015, kovtyukh_2022}. Given that the inner and outer disk stars generally represent different populations, it is unclear how this quantifies the effects of radial migration. We further investigate these populations in the rest of this section.

It is useful to consider the age-metallicity relation split into the two $\alpha$ sequences. The low-$\alpha$ sequence of young stars is prevalent in the midplane of the disk, particularly near the Sun and at larger radii $R$, whereas the older high-$\alpha$ sequence is predominantly in the central part and at large vertical distances $|z|$ (see Section~\ref{subsec:age-metallicity_discuss}). We confirm that the two sequences are indeed distributed this way by plotting the age-metallicity distribution of stars in our \emph{updated} sample in 12 disk zones as shown in Figure \ref{fig:age-metallicity-naive}. The blue and orange points correspond to the high- and low-$\alpha$ sequences respectively, which we obtain using the split from Figure \ref{fig:split_copula}. We use these data points to provide a naive estimate of the age-metallicity correlation for the two sequences in each disk zone. If the high-$\alpha$ sequence formed simultaneously as a well-mixed gas disk in a radially compact thick (or heated) component near the centre followed by the low-$\alpha$ sequence in a thin and extended distribution, then we can explain the naive correlation estimates in Figure \ref{fig:age-metallicity-naive} as well as the learned estimates in Figures \ref{fig:age-metallicity-z=0} and (top panels) \ref{fig:age-metallicity-z=all}. 

The middle and bottom panels of Figure \ref{fig:age-metallicity-z=all} show the learned age-metallicity correlation of the high-$\alpha$ and low-$\alpha$ sequences respectively. For the high-$\alpha$ sequence, we observe a $-0.4$ correlation near the central parts of the disk $R \lesssim 8$ kpc, which follows the high-$\alpha$ stellar distributions and correlation estimates shown in Figure \ref{fig:age-metallicity-naive}. Similarly, the learned correlation for the low-$\alpha$ sequence is consistent with being zero across the disk. There is also an unexpected positive correlation near the Sun, where most of our data is concentrated. Although unexpected, this is a feature of the data as seen in the naive estimates of Figure \ref{fig:age-metallicity-naive}. The solar neighbourhood zone $7 < R < 9$ kpc, $|z| < 0.5$ kpc in Figure \ref{fig:age-metallicity-naive} demonstrates that the metal-poor stars belonging to the high-$\alpha$ sequence are the oldest. On the other hand, the metal-rich stars have intermediate ages and solar metallicity stars are the youngest, both residing in the low-$\alpha$ sequence. These metal-rich intermediate age stars that migrated from the inner parts of the Galaxy lead to the slight positive correlation observed in the bottom left panel of \ref{fig:age-metallicity-z=all}. The same applies to the rest of the bottom panels with increasing $|z|$. 

Note that our sample size reduces with increasing $|z|$ and particularly for $|z| = 2$ kpc, the edge-case of our sample; therefore, we expect the rightmost panels in Figure \ref{fig:age-metallicity-z=all} to have the largest uncertainty. We also expect higher uncertainties in the central and outer parts of the disk due to smaller data samples (selection effects away from the Sun).

In order to verify our results using a larger sample, we repeat the estimation of $\rho(\tau, \M | R, z=0)$ for the \emph{original} sample we use for splitting the high- and low-$\alpha$ sequences. We observe the same trends as those discussed above, except those for the high-$\alpha$ sequence in the middle panels of Figure \ref{fig:age-metallicity-z=all}. This is because the ages in the \emph{original} sample come from the \texttt{astroNN} catalog whose precision is lower than that of \cite{leung_2023}. Particularly, the ages of the older stars are underestimated which results in dilution of the age-metallicity relation the high-$\alpha$ sequence.

\subsection{Constraints on radial migration models}
In the previous section, we qualitatively examine radial migration effects on the age-metallicity distribution across the Galactic disk. We now quantitatively constrain radial migration in the disk by comparing our estimate of stellar age-metallicity correlation with that of a mock data set sampled from the model in \cite{frankel_2020}. 

This model describes the chemodynamical evolution of the low-$\alpha$ disk of the Milky Way. It is built using the following ingredients: (1) stars are born on nearly circular orbits following an inside-out growth history; (2) there is a negative metallicity gradient with radius that evolves as stars move and chemically enrich the gas; (3) in-plane movement of stars is driven by diffusion in angular momentum and mean increase in radial action, i.e., radial migration (churning) and heating of stellar orbits (blurring) respectively; and (4) vertical heating results in increase of scale-height with age as well as flaring \citep{ting_rix_2019}. These model ingredients are defined by some parameters that are fit using a sample of APOGEE red clump stars. By sampling this best fit model, we can generate a mock low-$\alpha$ Galactic disk, and compare the spatial distribution of its stellar age-metallicity correlation with that in the bottom panels of Figure \ref{fig:age-metallicity-z=all}.

In order to properly compare the model with our estimates, we sample the mock low-$\alpha$ disk in such a way that its stellar distribution in $R, z$ is similar to that of our \emph{updated} APOGEE sample. For example, within a certain range of $R$ and $z$, both the mock and real (APOGEE) data samples will have the same number of stars. We then estimate the naive age-metallicity correlation of the mock sample in several disk zones binned in $R$ and $|z|$ (similar to that in Figure \ref{fig:age-metallicity-naive}). To compute an uncertainty on these model estimates, we repeat the sampling of the mock low-$\alpha$ disk $50$ times, and obtain the mean and variance of the binned correlation values. This variance is used to provide a $68$\% (one standard deviation) confidence interval on the model estimate. 

We show the comparison between our estimate and those from the mock data in Figure \ref{fig:frankel_compare}. The four panels show our learned functions $\rho(\tau, \M \mid R, |z|) \; \mathrm{for} \; |z| \in \{0, 0.5, 1, 1.5\}$ kpc, and the naive correlation estimates in bins of $|z| < 0.25$, $0.25 < |z| < 0.75$, $0.75 < |z| < 1.25$, and, $1.25 < |z| < 1.75$ kpc and $\Delta R = 1$ kpc. Note that we only investigate Galactocentric radii $5 < R < 15$ kpc. The model in \cite{frankel_2020} is fit to data within this range, so any predictions beyond that are an extrapolation. The conditional correlation functions are displayed as orange lines and labelled as ``data" since we estimate them by applying copulas and elicitable maps to APOGEE data. On the other hand, we represent the binned correlations using red dashed lines and name them ``model, present $R$" since they come from the mock low-$\alpha$ disk. Thus, comparing these ``data" and ``model" estimates allows us to understand how the model reproduces the evolution of the low-$\alpha$ disk. In particular, this test should help diagnose the effects of radial migration since we are looking at the age-metallicity correlation across the disk using the most statistically sound procedure developed till date. We note however that if by essence orbit evolution is driven by scatter, then using the variance in age-metallicity rather than the correlation would be a more direct evaluation.

To further look at radial migration effects, we replace the present radii of the stars in the mock sample with their birth radii and plot their age-metallicity correlation in bins of $R_\mathrm{birth}$ and $|z|$ using green dash-dotted lines. These ``model, birth $R$" estimates show that stellar age and metallicity are strongly correlated ($\rho \sim -0.8$) if stars do not move across radii in the Galactic disk. This is consistent with our expectation using one-zone models of evolution. However, after including radial migration, the ``model, present $R$" correlations are closer to zero and resemble the results we obtain in this paper. Thus, radial migration is in fact a dominant secular evolutionary process in the low-$\alpha$ disk. 

There are several reasons why the ``model" and ``data" are not perfectly consistent with each other, which overall indicate that the model presented in \cite{frankel_2020} is too simple to fully describe the data and their spatial variations. In particular, the model assumed a similar chemical enrichment rate at all Galactocentric radii, with a simple metallicity gradient. Additionally, the model assumed no spatial variations in radial migration efficiency. The differences found in this work could serve as a basis to diagnose where galactic modelling fails, and further improve Galactic models. Also, in this work, the model does not directly incorporate uncertainties in ages and metallicities of stars. The uncertainty in the model estimates is high towards the outer parts of the disk ($R > 12$ kpc) because stellar data is sparse in this region, whereas that near the Sun ($5 < R < 10$) is low due to more availability of data. However, as we go towards to the centre the Galaxy, we have greater contribution from older stars that have generally higher uncertainties on ages. Note that \cite{frankel_2020} account for these uncertainties when fitting the model to data, but the model itself does not provide an uncertainty on the age-metallicity correlation beyond sampling effects. We would like to add these uncertainties and see how they affect the comparison between model and data in the future.

Another reason for inconsistency is that the model does not explicitly include the effects of the bar and the high-$\alpha$ disk, and it is fit to a restricted sample of APOGEE red clump stars, whose intrinsic age distribution (peaking around 2 Gyr) makes them a poor tracer of the early history of the disk. Therefore, it is not an apples to apples comparison, and we should expect there to be discrepancies. Further, the model was fit to a dataset with a manual split of the low-$\alpha$ red clump stars, and additional cuts in observed stellar age. Any difference between the \cite{frankel_2020} and our split of $\alpha$ sequences will lead to subtle differences in results. We have significant contributions from metal-rich intermediate age and solar metallicity young red giant stars in our low-$\alpha$ sequence, which results in the slight positive correlation in the solar neighbourhood (as discussed in Section \ref{subsec:age-metallicity_discuss}). This effect is not seen in the model estimates, which could be due to variations in the low-$\alpha$ sequence split. However, in Figure \ref{fig:splits_compare}, we see that the \cite{frankel_2020} split is fairly consistent with ours, and so the positive correlation could be due to a different reason. Here, we emphasize the importance of developing automated approaches to split $\alpha$ sequences (as done in this paper), and using a consistent split across studies.

In the future, we would like to use a more comprehensive numerical model that is fit using a larger sample of red giant stars (which we use in this study). However, we note that the explicit modeling of radial migration in \cite{frankel_2020} is good for separately deducing its efficiency.

\section{Conclusions}\label{sec:conclusion}
The advent of high-resolution spectroscopic surveys has created opportunities to estimate the chemical and age distributions of the Milky Way with unprecedented accuracy and precision. The APOGEE survey used in this study densely observes spectra of red giants over a large portion of the Galactic disk. The spectra thus allow us to estimate abundances and ages of stars across the disk, which can then be decoded to understand the formation and evolution of the Galaxy.

The star formation history and chemical evolution of the Milky Way disk is governed by an interplay of several internal and external processes. We introduce two novel statistical methods called copulas and elicitable maps that have the potential to extract these processes using APOGEE data. In particular, we apply these powerful tools to investigate the secular process of radial migration, and its role in shaping the chemical and age-chemical distribution of stars in the Galactic disk. First, we apply copulas to the $[\alpha\mathrm{/Fe}]-[\mathrm{Fe/H}]$ abundance space of stars in the disk. Empirical studies have shown that this space has a bimodal structure representing the high- and low-$\alpha$ sequences that were traditionally defined as the geometric thin and thick disks \citep[e.g.,][]{anders_2014}. However, the separation between the two sequences is inconsistent across studies because it is often performed manually \citep[see][]{weinberg_2019}. This inconsistency makes studies of the individual sequences biased, and makes it difficult to determine if the two sequences truly exist \citep[as debated in e.g.,][]{rojas_2014} and whether they are distinct populations \citep[][]{mackereth_2017}. We resolve this issue by using copulas, statistical tools that examine the dependence structure of random variables without being affected by their marginal distributions. Copulas help us understand the true underlying relationship between $[\alpha\mathrm{/Fe}]$ and $[\mathrm{Fe/H}]$ independent of the star formation history of the disk. In addition, they provide non-linear measures of dependence and are invariant under increasing transformations, i.e., it does not matter if we use  linearly-scaled $Z/Z_\odot$ or the logarithmically-scaled $[\mathrm{Fe/H}]$ to describe metallicity. 

We find that the high- and low-$\alpha$ disk sequences have a clean separation in copula space as opposed to the data space, and therefore, we present the first data-driven and automated separation of $\alpha$ sequences in the Galactic disk using copulas. This separation reveals that the high-$\alpha$ disk ends at the same $[\alpha/\mathrm{Fe}]$ at high $[\mathrm{Fe/H}]$ as the low-$[\mathrm{Fe/H}]$ end of the low-$\alpha$ disk. Using the precise ages from \citet{leung_2023} further reveals that the sequences are also well separated in age. These findings support a sequential formation scenario for the high- and low-$\alpha$ disks.

By combining copulas with elicitable maps, we estimate the stellar age-metallicity (or age-$[\mathrm{Fe/H}]$) correlation as a continuous function of radius $R$ and distance from the midplane $z$ in the disk. A statistical mapping or function is said to be elicitable if one can estimate it by minimizing a loss or \textit{scoring function}. Thus, elicitable maps allow us to obtain continuous functions using sparse data; we apply them to APOGEE data to estimate the age-metallicity correlation conditional on radius $R$ and height $z$ in the disk, i.e., $\rho(\tau, \M | R, z)$. The correlation is extracted using the copula of age and metallicity, and the elicitable map estimation is performed using neural networks that minimize the relevant loss function. The $\rho(\tau, \M | R, z)$ estimate encompasses the range $0 < R < 20$ kpc and $|z| < 2$ kpc, and provides avenues to assess the spatial distribution of the age-metallicity relation beyond the broad brush binned estimates in the literature. 

We find that age and metallicity are uncorrelated in the midplane from $5\lesssim < R < 20\,\mathrm{kpc}$ and that they display a negative correlation of $\approx -0.4$ in the bar region at $R \lesssim 5\,\mathrm{kpc}$. At larger $z$, the disk has a  $\approx -0.4$ age--metallicity correlation over a larger radial range. Splitting the sample into high- and low-$\alpha$ reveals that this is driven by the fact that the high-$\alpha$ sequence has a negative correlation wherever it is present, while the low-$\alpha$ disk has zero correlation over the entire radial range of the disk. These findings are all consistent with the scenario where spiral-driven radial migration removes the initial age--metallicity correlation for stars on kinematically-cold orbits (primarily in the low-$\alpha$ disk outside the bar). We also compare our age--metallicity results with the best fit radial-migration models in \citet{frankel_2020}, finding good agreement and, thus, strengthening the case that radial migration is strong in the Milky Way disk and has strong effects on its age and chemical distribution.

We note that the accuracy and precision of our results are heavily dependent on that of the abundance and age estimates and that the precision is limited by the relatively-small number of stars in our sample. In the future, we would like to use the Milky Way Mapper of the ongoing SDSS-V survey and apply novel statistical methodology \citep[e.g.,][]{patil_2022a} to push the boundaries of current abundance catalogues in terms of quality, quantity, and scope. To further improve age estimates, we would like to expand the variational encoder-decoder model in \citep{leung_2023} to larger samples of disk stars. The model is trained using asteroseismic ages of stars in the Kepler survey, which could be applied to more expansive surveys such as TESS (and the upcoming PLATO mission) using robust frequency analysis methods \citep[e.g.,][]{patil_2022b}.

In conclusion, copulas and elicitable maps are powerful statistical tools that can address several questions on the formation and evolution of the Galactic disk as well as other data-rich problems in astronomy.

\section*{Acknowledgements}
AAP and this project are supported by the Data Sciences Institute at the University of Toronto. AAP, JB, and HWL acknowledge financial support from NSERC (funding reference number RGPIN-2020-04712). NF is supported by the Natural Sciences and Engineering Research Council of Canada (NSERC), funding reference number CITA 490888-16, through a CITA postdoctoral fellowship, and acknowledges partial support from an Arts \& Sciences Postdoctoral Fellowship at the University of Toronto. 

The authors thank Gwendolyn Eadie for insightful feedback during the early stages of this work. 

This work has made use of data from the European Space Agency (ESA) mission {\it Gaia} (\url{https://www.cosmos.esa.int/gaia}), processed by the {\it Gaia} Data Processing and Analysis Consortium (DPAC, \url{https://www.cosmos.esa.int/web/gaia/dpac/consortium}). Funding for the DPAC has been provided by national institutions, in particular the institutions participating in the {\it Gaia} Multilateral Agreement. Funding for the Sloan Digital Sky Survey IV has been provided by the Alfred P. Sloan Foundation, the U.S. Department of Energy Office of Science, and the Participating Institutions. SDSS-IV acknowledges support and resources from the Center for High-Performance Computing at the University of Utah. The SDSS web site is \url{www.sdss.org}.


\section*{Data Availability}
The SDSS-IV/APOGEE DR17 data described in Section \ref{sec:data} is publicly available. In particular, the \texttt{astroNN} abundances, distances, and ages are publicly released as a value-added catalog of DR17. The variational encoder-decoder ages in \cite{leung_2023} are available in the GitHub repository at \url{https://github.com/henrysky/astroNN_ages}. We also provide all the code developed for this article at \url{https://github.com/aaryapatil/elicit-disk-copulas}.











\bsp	
\label{lastpage}
\end{document}